\documentclass[useAMS,usenatbib]{mn2e}
\usepackage{amsmath,amssymb}
\usepackage{graphicx,mathptm}
\usepackage{times}
\usepackage{natbib}
\usepackage{psfig}

\title[BH scaling relations and the AGN luminosity function]
{Modeling the cosmological co-evolution of supermassive black holes and galaxies: 
I. BH scaling relations and the AGN luminosity function}

\author[Marulli et al.]
{
Federico Marulli$^1$, 
Silvia Bonoli$^2$,
Enzo Branchini$^3$,
Lauro Moscardini$^{1,4}$,
Volker Springel$^2$
\\ 
$^1$Dipartimento di Astronomia, Universit\`a degli Studi di Bologna, 
via Ranzani 1, I-40127 Bologna, Italy \\
$^2$Max-Planck-Institut f\"ur Astrophysik, Karl-Schwarzschild Strasse 1,
D-85740 Garching, Germany\\
$^3$Dipartimento di Fisica, Universit\`a 
degli Studi ``Roma Tre'', via della Vasca Navale 84, I-00146 Roma, Italy\\
$^4$INFN, Sezione di Bologna, viale Berti Pichat 6/2, I-40127 Bologna, Italy\\
}

\begin{document}

\maketitle

\label{firstpage}

\begin {abstract}
We model the cosmological co-evolution of galaxies and their central
supermassive black holes (BHs) within a semi-analytical framework
developed on the outputs of the Millennium Simulation. This model,
described in detail in \citet{croton2006} and \citet{delucia2007},
introduces a `radio mode' feedback from Active Galactic Nuclei (AGN) at
the centre of X-ray emitting atmospheres in galaxy groups and
clusters. Thanks to this mechanism, the model can simultaneously
explain: (i) the low observed mass drop-out rate in cooling flows; (ii)
the exponential cut-off in the bright end of the galaxy luminosity
function; and (iii) the bulge-dominated morphologies and old stellar
ages of the most massive galaxies in clusters.  This paper is the first
of a series in which we investigate how well this model can also
reproduce the physical properties of BHs and AGN.  Here we analyze the
scaling relations, the fundamental plane and the mass function of BHs,
and compare them with the most recent observational data.  Moreover, we
extend the semi-analytic model to follow the evolution of the BH mass
accretion and its conversion into radiation, and compare the derived AGN
bolometric luminosity function with the observed one.  While we find for
the most part a very good agreement between predicted and observed BH
properties, the semi-analytic model underestimates the number density of
luminous AGN at high redshifts, independently of the adopted Eddington
factor and accretion efficiency. However, an agreement with the
observations is possible within the framework of our model, provided it
is assumed that the cold gas fraction accreted by BHs at high redshifts
is larger than at low redshifts.
\end {abstract}

\section {Introduction}

Over the last years, several observations have demonstrated that
supermassive black holes likely reside at the centres of all spheroidal
galaxies \citep[see e.g.][]{kormendy1995,richstone1998}.  Even more
interestingly, their properties appear to strongly correlate with those
of their host galaxies \citep{magorrian1998, ferrarese2000,
gebhardt2000, graham2001, tremaine2002, mclure2002, baes2003,
marconi2004, haring2004, feoli2007, graham2007} and, apparently, also
with the ones of the whole host dark matter (DM) haloes
\citep{ferrarese2002, baes2003, ferrarese2005}. Although it is not yet
clear which of these relations is ``more fundamental'' \citep[see
e.g.][]{novak2006}, they reasonably suggest a close link between the
assembly history of the BHs and the cosmological evolution of galaxies.
Most recently, \citet{hopkins2007a} have shown that these relationships
are not independent and could be interpreted as different projections of
a BH fundamental plane, analogous to the fundamental plane of
elliptical galaxies.  The striking similarity between these two
fundamental planes is another clue that galaxy spheroids and BHs do not
form and evolve independently.

The paradigm that AGN are powered by mass accretion onto 
BHs \citep{salpeter1964, lynden1969} has also received very strong
support from spectroscopic and photometric observations of the stellar
and gas dynamics in the central regions of local spheroidal galaxies and
bulges. Moreover, by estimating the total energy radiated by AGN during
their whole life, it can be shown that nearly all the mass in BHs has
been accumulated during periods of bright AGN activity
\citep{soltan1982}, implying that the common physical process which
produces galaxy spheroids and BHs must also be responsible for
triggering bright AGN.

Such a cosmological co-evolution of BHs, AGN and galaxies is expected
in the standard framework, in which cosmic structures grow
hierarchically via gravitational instability and merging events
destabilize the gas at the galaxy centres, triggering star formation and
BH mass accretion.  In order to investigate this complex scenario,
several models have been developed, based on either pure analytic
approximations \citep[see, e.g.,][]{efstathiou1988, haehnelt1993,
haiman1998, percival1999, haiman2000, martini2001, wyithe2003,
hatziminaoglou2003, hopkins2007b}, or semi-analytic ones \citep[see,
e.g.,][]{kauffmann2000, cavaliere2002, enoki2003, volonteri2003a, granato2004, 
springel2005, cattaneo2005, croton2006,  malbon2006, fontanot2006}.
Recently, thanks to the availability of unprecedented computational
power, fully numerical models have also become available \citep[see,
e.g.,][]{dimatteo2005,springel2005c,li2007,sijacki2007,dimatteo2007}.
 
Simple analytic models in which AGN activity is only triggered by DM
halo major mergers succeeded in quantitatively describing the observed
evolution of the AGN number counts and luminosity at all but low
redshifts, provided that some mechanism is advocated to inhibit
accretion within massive haloes hosting bright AGN. However they fail
in reproducing the observed AGN clustering at high redshifts
\citep{marulli2006}.  Slightly more sophisticated semi-analytic models
in which the halo merger history and associated BHs are followed by
Monte Carlo realizations of the merger hierarchy, while the baryonic
physics is neglected as well, can correctly reproduce both the AGN
luminosity and clustering function at $z\gtrsim1$ \citep{marulli2006},
but the number density of faint AGN is significantly below
observations, a clear indication that DM halo mergers cannot constitute
the only trigger to accretion episodes in the local BH population
\citep{marulli2007}, and that in order to properly describe the
cosmological evolution of BHs and AGN, the main baryonic phenomena
involving the gas contents of DM halos cannot be neglected.

This complication is reminiscent of the one found in the description of
galaxies, where the well-known mismatch in shape between the predicted
distribution of DM halo masses and the observed distribution of galaxy
luminosities requires the consideration of complex baryonic phenomena
like, for instance, cooling inefficiencies to reduce gas condensation in
massive structures, or supernova \citep{white1978, white1991} and
stellar kinetic feedback \citep{fontanot2006} to remove cold gas in low
mass systems, as well as photoionisation heating to suppress the
formation of dwarfs \citep{efstathiou1992}.  Cooling effects alone are
however too weak to produce the bright end cut-off of the luminosity
function, and it seems to be mandatory to include additional feedback
processes in massive halos
\citep[e.g.][]{benson2003,fontanot2006,croton2006,ciotti2007}.  Standard models of
galaxy formation face two additional problems: i) the persistence of a hot gas
atmosphere at the centre of most galaxy clusters despite the fact that
the local cooling time is much shorter than the age of the system
\citep[see, e.g.][and references therein] {cowie1977, fabian1977,
peterson2001, tamura2001, fabian2003, mcnamara2005, morandi2007}, and
ii) the fact that most massive galaxies, typically ellipticals in
clusters, are made of the oldest stars and so finished their star
formation earlier than lower mass galaxies \citep[see, e.g.][and
references therein]{cowie1996, delucia2006, cimatti2006}.

In this paper we study the cosmological co-evolution of galaxies and
their central BHs using a semi-analytical model developed on the outputs
of the Millennium Simulation and described in detail in
\citet{croton2006} and \citet{delucia2007}.  In this scenario, {\em
radio mode} feedback from AGN at the centre of galaxy groups and
clusters is invoked to prevent significant gas cooling in large halos,
thus limiting the mass of the central galaxies and preventing them from
forming stars at late times when their mass and morphology can still
change through mergers. Thanks to this mechanism, \citet{croton2006}
demonstrated that such a model can simultaneously explain the low
observed mass drop-out rate in cooling flows, the exponential cut-off in
the bright-end of the galaxy luminosity function, and the
bulge-dominated morphologies and stellar ages of the most massive
galaxies in clusters.  
\\

Here we are interested in investigating how well
this model can also reproduce the statistical properties of BHs and
AGN.  To do that, we extend the original model by adding new
semi-analytical prescriptions to describe the BH mass accretion rate in
the accretion episodes triggered by galaxy mergers, which fuel the {\em
quasar mode}, and their conversion into radiation. We then analyze the
scaling relations, the fundamental plane and the mass function of BHs,
and compare them with the most recent observational data available. Finally, we compare the predicted AGN bolometric luminosity
function with the observed one, and propose some modifications to the
original semi-analytic assumptions to better fit the data.
\\

The paper is organized as follows. In Section \ref{sec:description}, we
briefly describe the main aspects of our semi-analytic model and
illustrate the new equation introduced to describe the BH mass accretion
in the {\em quasar mode} in more detail.  In Section
\ref{sec:comparison}, we compare the model predictions with the best
observational data available for the BH and AGN populations. Finally, in
Section \ref{sec:conclusions} we summarize our conclusions.

\section {The model} \label{sec:description}

Our semi-analytic model for the co-evolution of DM haloes, galaxies and
their central BHs consists of three ingredients, that we describe
separately in this section: a numerical simulation to obtain the merger
history of the DM haloes, a set of analytic prescriptions to trace the
evolution of galaxies within their host haloes and a set of recipes to
follow the BH accretion history and the AGN phenomenon.

\subsection{Numerical simulation}

In this work we use the outputs of the Millennium Simulation, which
followed the dynamical evolution of $2160^3\simeq 10^{10}$ DM particles
with mass $8.6\times10^8\,h^{-1}{\rm M}_{\odot}$ in a periodic box of
$500\,h^{-1}$Mpc on a side, in a $\Lambda$CDM ``concordance"
cosmological framework \citep{springel2005}.  The computational box is
large enough to include rare objects such as quasars or rich galaxy
clusters, the largest of which contain about 3 million simulation
particles at $z\!=\!0$. At the same time, the mass resolution is high
enough to resolve the DM halo of $0.1\,L_\star$ galaxies with
$\sim\!100$ particles.  The short-range gravitational force law is
softened on the co-moving scale $5\,h^{-1}{\rm kpc}$
(Plummer-equivalent) which may be taken as the spatial resolution limit
of the calculation.  The cosmological parameters (the matter density
parameter $\Omega_{\rm m}=0.25$, the baryon density parameter
$\Omega_{\rm b}=0.045$, the Hubble parameter $h=H_0/100\, {\rm km\,
s^{-1} Mpc^{-1}}=0.73$, the cosmological constant contribution to the
density parameter $\Omega_\Lambda=0.75$, the primordial spectral index
$n=1$, and the power spectrum normalization $\sigma_8=0.9$), are
consistent with determinations from the combined analysis of the
2-degree Field Galaxy Redshift Survey (2dFGRS) \citep{colless2001} and
first-year WMAP data \citep{spergel2003}, as shown by
\cite{Sanchez2006}.  We recall that the more recent analysis of the WMAP
3-year data \citep{spergel2007} suggests slightly different values (in
particular smaller values for $\Omega_{\rm m}$, $\sigma_8$ and
$n$). However, as demonstrated by \citet{wang2007}, due to the current
modelling uncertainties, it is not possible to distinguish the two WMAP
cosmologies on the basis of the observed galaxy properties, since the
variations induced by acceptable modifications of the free parameters of
the galaxy formation model are at least as large as those produced by
the variation in the cosmological parameters.

The Millennium Simulation was carried out with a special version of the
{\small GADGET-2} code \citep{springel2005gadget2}, optimized for very
low memory consumption, at the Computing Centre of the Max-Planck
Society in Garching, Germany.  We make use of hierarchical merging trees
extracted from this simulation which encode the full formation history
of DM haloes and subhalos, previously identified with, respectively, a
friends-of-friends (FOF) group-finder and an extended version of the
{\small SUBFIND} algorithm \citep{springel2001b}.  These trees
constitute the backbone of our semi-analytic model, which is implemented
during the post-processing phase: this allows us to simulate the wide
range of baryonic processes occurring during the formation and evolution
of galaxies and their central BHs.

\subsection{Galaxy evolution}

We use the galaxy formation model of \citet{croton2006} as updated by
\citet{delucia2007}. Although not in agreement with some properties of
the red and blue galaxy populations \citep[see,
e.g.,][]{weinmann2006,kitzbichler2007}, this model is able to reproduce
the overall observed properties of galaxies, i.e. the relations between
stellar mass, gas mass and metallicity, the luminosity, colour and
morphology distributions \citep{croton2006,delucia2006}, the two-point
galaxy correlation functions \citep{springel2005}, and the global galaxy
luminosity and mass functions at high redshift \citep{kitzbichler2007}.
We refer to the original papers for a full description of the numerical
implementation of the model.  In the following, we briefly recall the
treatment of the physical processes involved in the galaxy evolution,
and describe the prescriptions for the BH growth and the AGN evolution.

Following the standard paradigm set out by \citet{white1991} and adapted
to high-resolution N-body simulations by \citet{springel2001b}, we
assume that as a DM halo collapses, a fraction $f_b=0.17$ of its mass is
in the form of baryons and collapses with it, consistent with the
first-year WMAP result \citep{spergel2003}. Initially, these baryons are
in the form of a diffuse gas with primordial composition, but later they
include gas in several phases as well as stars and heavy elements.
Conventionally, with the simplifying assumption of an ideal gas which
cools isobarically, the cooling time of the gas is computed as the ratio
of its specific thermal energy to the cooling rate per unit volume,
\begin{equation}
t_{\rm cool} = \frac{3}{2} \frac{\bar{\mu} m_p k T }{ \rho_g (r) \Lambda
(T,Z)} ~,
\label{tcool}
\end{equation}
where $\bar{\mu} m_p$ is the mean particle mass, $k$ is the Boltzmann
constant, $\rho_g(r)$ is the hot gas density, and $\Lambda (T,Z)$ is the
cooling function \citep{sutherland1993,maio2007}.  Equation
(\ref{tcool}) is valid at temperature higher than $\sim10^4$ K, where
hydrogen and helium remain ionized and the number of particles remains
approximately constant.

We assume the post-shock temperature of the infalling gas to be the
virial temperature of the halo, $T=35.9\,(V_{\rm
vir}/\rm{km\,s^{-1}})^2$ K, where $V_{\rm vir}$ is the halo virial
velocity. Moreover, we assume that the hot gas within a static
atmosphere has a simple `isothermal' distribution,
\begin{equation}
\rho_g(r) = \frac{m_{\rm hot}}{4 \pi R_{\rm vir} r^2} ~,
\label{rhog}
\end{equation}
where $m_{\rm hot}$ is the total hot gas mass associated with the halo
and is assumed to extend to its virial radius $R_{\rm vir}$.

In order to estimate an instantaneous cooling rate onto the central
object of a halo, given its current hot gas content, we define the
cooling radius, $r_{\rm cool}$, as the radius at which the local cooling
time (assuming the structure of equation (\ref{rhog})) is equal to the
halo dynamical time, $R_{\rm vir}/V_{\rm vir}= 0.1\, H(z)^{-1}$
\citep{springel2001,delucia2004,croton2006}; here $H(z)$ represents the
redshift evolution of the Hubble constant.  The cooling rate can then be
determined through the following continuity equation,
\begin{equation}
\dot{m}_{\rm cool} = 4 \pi \rho_g(r_{\rm cool}) r_{\rm cool}^2
\dot{r}_{\rm cool} ~. 
\label{mdotcool}
\end{equation}
More details about our cooling prescriptions can be found in
\citet{croton2006}.
 
The photo-ionization heating of the intergalactic medium suppresses the
concentration of baryons in shallow potentials \citep{efstathiou1992},
and can be responsible of the inefficient accretion and cooling in
low-mass haloes.  Following \citet{gnedin2000}, we model the effect of
such photo-ionization heating by defining a characteristic mass scale,
$M_{\rm F}$, below which the gas fraction $f_b$ is reduced relatively to
the universal value:
\begin{equation}
f_{\rm b}^{\rm halo}(z,M_{\rm vir}) = \frac{f_{\rm b}^{\rm cosmic}}{[1 +
  0.26 \,M_{\rm F}(z) / M_{\rm vir}]^3}~.
\label{reion}
\end{equation}
We adopt the $M_{\rm F}(z)$ parameterization of \citet{Kravtsov2004},
which results in a filtering mass $M_{\rm F}$ of $4\times 10^{9}
M_{\odot}$ at the reionization epoch, and $3 \times 10^{10}M_{\odot}$ by
the present day \citep[but see][]{hoeft2006}.

In the semi-analytic framework we use in this work, the star formation
is assumed to occur at a rate given by:
\begin{equation}
\dot{m}_{*}=\alpha_{\rm SF}(m_{\rm cold}- m_{\rm crit})/t_{\rm dyn,disc} ~,
\label{m*}
\end{equation}
where $m_{\rm cold}$ is the cold gas mass, $t_{\rm dyn,disc}$ is the
dynamical time of the galaxy, defined as the ratio between the disk
radius and the virial velocity, $m_{\rm crit}$ corresponds to a critical
value for the gas surface density
\citep{kauffmann1996,kennicutt1998,mo1998}, and $\alpha_{\rm SF}=0.03$
controls the efficiency of the transformation of cold gas into stars.
Massive stars explode as supernovae shortly after star formation events
and are assumed to reheat a gas mass proportional to the mass of stars:
\begin{equation}
{\Delta}m_{\rm reheated} = \epsilon_{\rm disk}{\Delta}m_{*},
\end{equation}
where we set the free parameter $\epsilon_{\rm disk}=3.5$ based on the
observational data.  The energy released by an event which forms a mass
${\Delta}m_{*}$ in stars is assumed to be:
\begin{equation}
{\Delta}E_{\rm SN}=0.5\epsilon_{\rm halo}{\Delta}m_{*}V^2_{SN},
\end{equation}
where $0.5V^2_{\rm SN}$ is the mean supernova energy injected per unit
mass of newly formed stars, and $\epsilon_{\rm halo}$ represents the
efficiency with which this energy is able to convert cold interstellar
medium into hot, diffuse halo gas. The amount of gas that leaves the DM
halo in a ``super-wind'' is determined by computing whether excess SN
energy is available to drive the flow after reheating of material to the
halo virial temperature.

We model the disk instabilities using the analytic stability criterion
of \citet{mo1998}; the stellar disk of a galaxy becomes unstable when
the following inequality is met:
\begin{equation}
\frac{V_c}{({\rm G} m_{\rm{disk}} / r_{\rm{disk}})^{1/2}} \le 1 ~.
\label{disk_stability}
\end{equation}
At each time-step we evaluate the left-hand side of equation
(\ref{disk_stability}) for each galaxy, and if it is smaller than unity
we transfer enough stellar mass from disk to bulge (at fixed
$r_{\rm{D}}$) to restore stability.

In the Millennium Run, substructures are followed down to masses of
$1.7\times 10^{10}h^{-1}M_\odot$, so that we can properly follow the
motion of galaxies inside their hosting DM haloes until tidal truncation
and stripping disrupt their subhalos at this resolution limit. At this
point, we estimate a survival time for the galaxies using their current
orbit and the dynamical friction formula of \citet{binney1987}
multiplied by a factor of $2$, as in \citet{delucia2007}. After this
time, the galaxy is assumed to merge onto the central galaxy of its own
halo.  Galaxy mergers induce starburst which we describe using the
``collisional starburst'' prescription introduced by
\citet{somerville2001}.  In this model, a fraction $e_{\rm burst}$ of
the combined cold gas from the two merging galaxies is turned into stars
as follows:
\begin{equation}
e_{\rm burst} = \beta_{\rm burst} (m_{\rm sat} / m_{\rm
  central})^{\alpha_{\rm burst}} ~,
\label{e_burst}
\end{equation}
where the two parameters are taken as $\alpha_{\rm burst}=0.7$ and
$\beta_{\rm burst}=0.56$, appropriate for merger mass ratios ranging
from 1:10 to 1:1 \citep{cox2004}.

\subsection{BH mass accretion and AGN}

\subsubsection{The `radio mode'} \label{radio}

When a static hot halo has formed around a galaxy, we assume that a
fraction of the hot gas continuously accretes onto the central BH,
causing a low-energy `radio' activity in the galaxy centre.  Following
\citet{croton2006}, the BH mass accretion rate during these phases is
postulated to scale as follows:
\begin{equation}
\dot{M}_{\rm BH,R} = \kappa_{\rm{AGN}} \left(\frac{M_{\rm BH}}{10^{8}
  M_{\odot}}\right) \left(\frac{f_{\rm hot}}{0.1}\right)
  \left(\frac{V_{\rm vir}}{200\,\rm{km\,s^{-1}}}\right)^3 ~, 
\label{accretionR}
\end{equation}
where $M_{\rm BH}$ is the BH mass, $f_{\rm hot}$ is the fraction of the
total halo mass in the form of hot gas, and $\kappa_{\rm{AGN}}$ is a
free parameter set equal to $7.5\times10^{-6} \rm{M_{\odot} yr^{-1}}$ in
order to reproduce the turnover at the bright end of the galaxy
luminosity function.  Since $f_{\rm hot}$ is approximately constant for
$V_{\rm vir} \gtrsim 150\,{\rm km\,s^{-1}}$, the dependence of
$\dot{m}_{\rm BH,R}$ on this quantity has a little effect. Note that the
accretion rate given by equation (\ref{accretionR}) is typically
orders-of-magnitude below the Eddington limit. In fact, the total mass
growth of BHs in the radio relative to the quasar mode discussed below
is negligible.

It is also assumed that the {\em radio mode} feedback injects energy
efficiently into the surrounding medium, which can reduce or even stop
the cooling flow in the halo centres. The mechanical heating generated
by this kind of BH mass accretion and described as $L_{\rm BH} =
\epsilon\dot{M}_{\rm BH}c^2$, where $\epsilon = 0.1$ is the {\em
accretion efficiency} and $c$ is the speed of light, induces a modified
infall rate of the following kind:
\begin{equation}
\dot{m}_{\rm cool}' = \dot{m}_{\rm cool} - 
\frac{L_{\rm BH}}{0.5 V_{\rm vir}^2} ~.
\label{effective_cool}
\end{equation}
For consistency we never allow $\dot{m}_{\rm cool}'$ to fall below zero.
In this scenario, the effectiveness of radio AGN in suppressing cooling
flows is greatest at late times and for large values of the BH
mass, which is required to successfully reproduce the luminosities,
colours and clustering of low-redshift bright galaxies.

\subsubsection{The `quasar mode'} \label{quasar}

In our model BHs accrete mass after a galaxy merger both through
coalescence with another BH and by accreting cold gas, the latter being
the dominant accretion mechanism.  For simplicity, the BH coalescence is
modelled as a direct sum of the progenitor masses, thus ignoring
gravitational wave losses.  Following \citet{kauffmann2000}, we assume
that the gas mass accreted during a merger is proportional to the total
cold gas mass present, but with an efficiency which is lower for smaller
mass systems and for unequal mergers:
\begin{equation}
\Delta M_{\rm BH,Q} = \frac{f'_{\rm BH} \ m_{\rm cold}}{1 +
  (280\,\rm{km\,s^{-1}}/V_{\rm vir})^2}\, , 
\label{eq:accretionQ}
\end{equation}
where 
\begin{equation} 
f'_{\rm BH} = f_{\rm BH}\ (m_{\rm sat}/m_{\rm central})\, ,
\label{eq:fBH}
\end{equation}
and $f_{\rm BH}\approx 0.03$ is chosen to reproduce the observed local
$M_{\rm BH}-M_{\rm bulge}$ relation.  Thus, any merger-induced
perturbation to the gas disk (which might come from a bar instability or
a merger-induced starburst) can in principle drive gas onto the central
BH. However, the fractional contribution of minor mergers is typically
quite small, so that accretion driven by major mergers is the dominant
mode of BH growth in our scenario.  This kind of accretion, which we
call {\em quasar mode}, is also closely associated with starbursts,
which occur concurrently. We do not model feedback from the quasar
activity in the current model, but it can be approximately represented
by an enhanced effective feedback efficiency for the supernovae
associated with the intense starburst.

\subsubsection{AGN luminosity} \label{sec:new}

\begin{figure}
  \includegraphics[width=0.45\textwidth]{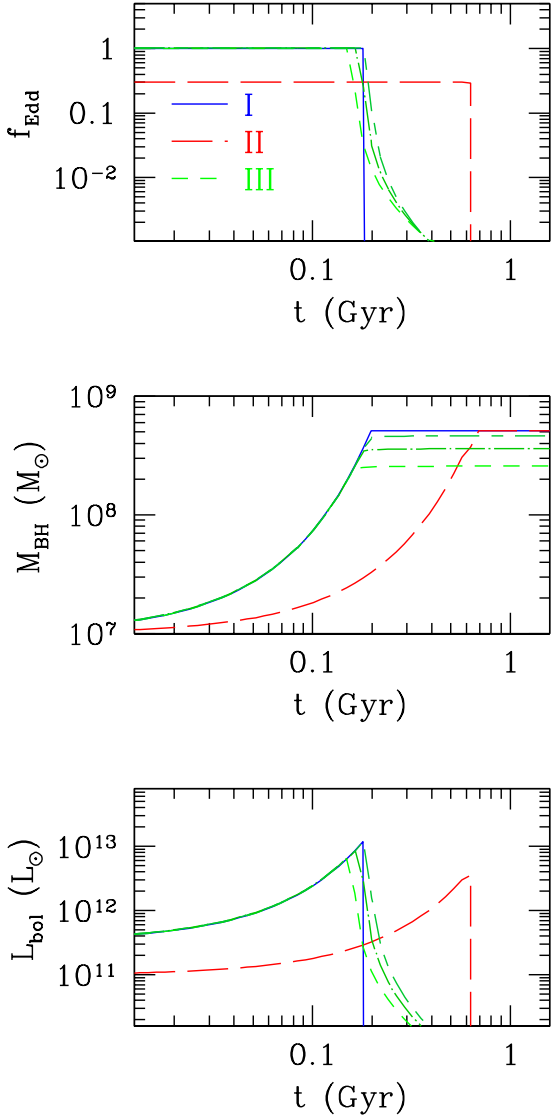} 
  \caption{ The time evolution of $f_{\rm Edd}$ (top panel), $M_{\rm
    BH}$ (central panel) and $L_{\rm bol}$ (bottom panel) for our three
    lightcurve models ({\em I} (blue solid lines), {\em II} (red
    short-dashed lines) and {\em III} (green lines)), for an
    illustrative case of a BH of mass $M_{\rm BH}=10^7M_\odot$ accreting
    a mass $\Delta M_{\rm BH,Q}=5\times10^8M_\odot$, starting at $z=3$.  The three green curves,
    showing our model {\em III}, have been obtained by setting $\mathcal{F}=0.5$ 
    (short dashed), $0.7$ (dotted-long dashed) and $0.9$
    (short dashed-long dashed).  }
  \label{fig:accr}
\end{figure}

The output of the model summarized hitherto, called {\em DeLucia2006a
catalogue} \citep{delucia2007}, is publicly available at
http://www.mpa-garching.mpg.de/millennium \citep{lemson2006}. In this
default model, for simplicity, the BH mass accretion triggered by each
merger is implemented as an instantaneous event and the BH seed masses
are set equal to zero.

In order to study the evolution of AGN inside this cosmological
framework, we improve the original model of \citet{delucia2007} by
adding new semi-analytical prescriptions to describe the BH mass
accretion rate during each merger event in the {\em quasar mode}, and
its conversion into radiation. In this implementation, BHs do not accrete mass instantaneously. 
Instead, the accretion is coupled to the light curve model adopted.
If a galaxy undergoes a merger while the central BH is still accreting mass from 
a previous merger, the cold gas still to be accreted is added to the new gas 
reservoir, and the accretion re-starts under the new physical conditions.
In Sect. \ref{sub:dephist}  we show that the BH scaling relations are weakly affected by this change. 
We use the following definitions to
parameterize the bolometric luminosity emitted by accretion onto BHs, as
a function of the {\em accretion efficiency}, $\epsilon$, and the {\em
Eddington factor}, $f_{\rm Edd}(t):=L_{\rm bol}(t)/L_{\rm Edd}(t)$,
\begin{eqnarray} 
  L_{\rm bol}(t) & = & \frac{\epsilon}{1-\epsilon}\dot{M}_{\rm BH}(t)c^2 \nonumber \\
  & = & f_{\rm Edd}(t)L_{\rm Edd}(t)=f_{\rm Edd}(t)\frac{M_{\rm BH}(t)}{t_{\rm Edd}}c^2,  \nonumber 
  \label{eq:Lagn}
\end{eqnarray} 
\begin{equation} 
  \Longrightarrow d\ln M_{\rm BH}(t) = \frac{dt}{t_{\rm ef}(t)}, 
  \label{eq:Mdot}
\end{equation} 
where $L_{\rm Edd}$ is the Eddington luminosity, 
$t_{\rm Edd}=\sigma_{T} c /(4\pi m_{p} G) \sim0.45\,{\rm Gyr}$ and $t_{\rm
ef}(t)=\frac{\epsilon}{1-\epsilon}\frac{t_{\rm Edd}}{f_{\rm Edd}(t)}$ is
the e-folding time ($t_{\rm ef}\equiv t_{\rm Salpeter}$ if $f_{\rm Edd}=1$).

No strong observational constraints are available for $\epsilon$ and
$f_{\rm Edd}$, the parameters that regulate the BHs powering the AGN
and, more importantly, if and how they depend on redshift, BH masses,
AGN luminosities and so on.  
However, some observations at $z\sim 0$ indicate
that $0.04<\epsilon<0.16$ and  $0.1<f_{\rm Edd}<1.7$ \citep{marconi2004}.
Furthermore, it has been suggested that $f_{\rm Edd}$ may depends
on redshift \citep{shankar2004} and BH mass \citep{netzer2007}.
In this paper, for simplicity, we do not follow the evolution of the BH spins
\citep[see, e.g.][and references therein]{volonteri2007} and we take a
constant mean value for the accretion efficiency of
$\epsilon=\left<\epsilon\right>=0.1$ at all redshifts.

\begin{figure*}
  \includegraphics[width=1.0\textwidth]{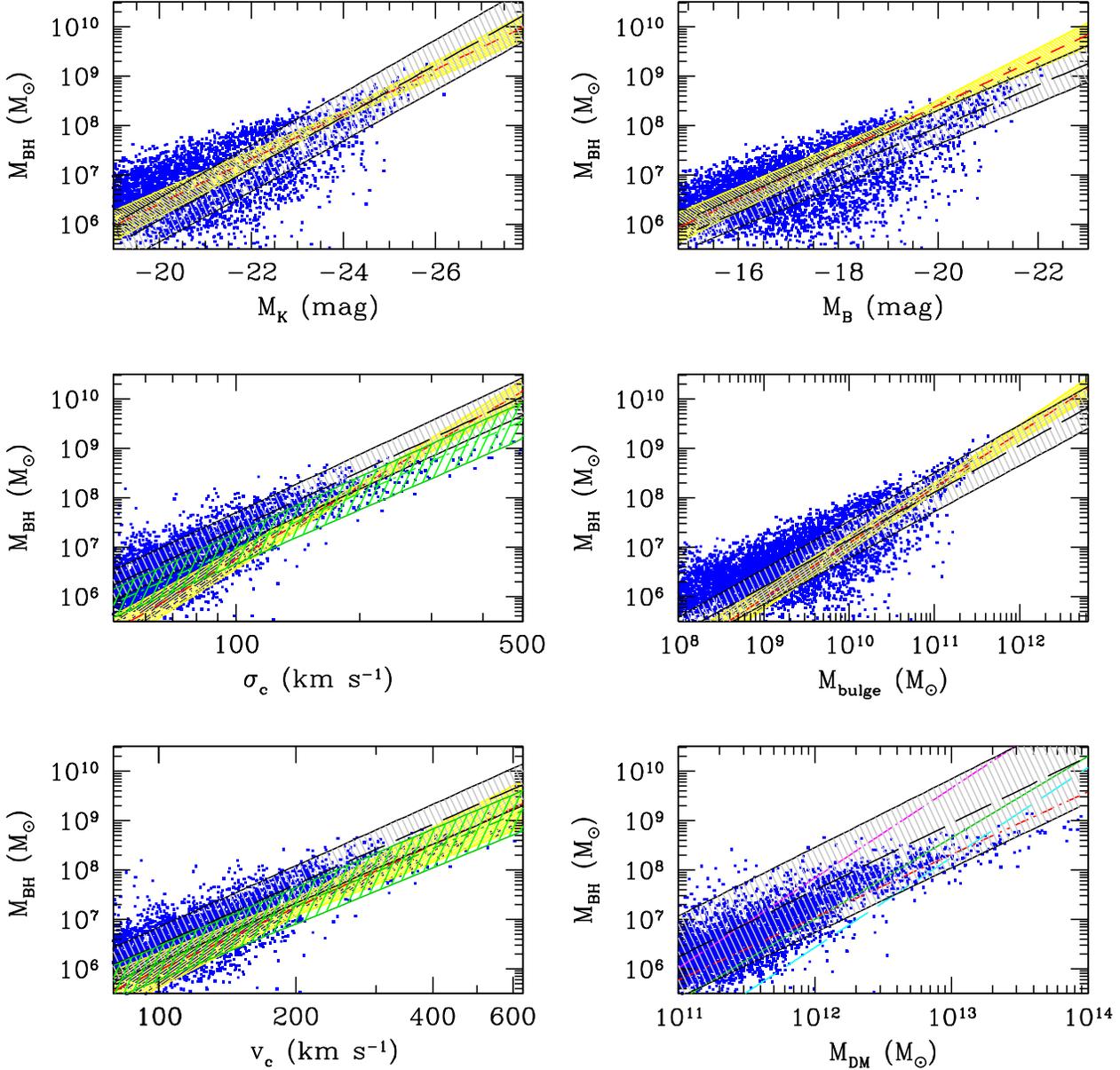} 
\caption{
  Starting from the upper left panel down to the bottom right one,
  scaling relations between the masses of the central BHs in the
  simulated galaxies with six different properties of their hosts: the
  K- and B-band bulge magnitude (top left and right panels,
  respectively), the bulge velocity dispersion and mass (central left
  and right panels, respectively), the circular velocity of the galaxy
  (bottom left panel) and the virial mass of the DM halo (bottom right
  panel). Blue dots represent the outputs of the {\em DeLucia2006a
  catalogue}, grey and yellow shaded areas show the best fit to the
  model predictions and to the observational datasets, respectively.
  Starting from the upper left panel down to the lower right, the yellow
  shaded areas refer to the best-fit relations obtained by
  \citet{marconi2004} (the upper two panels of the plot),
  \citet{ferrarese2005}, \citet{haring2004}, \citet{baes2003} and, in
  the lower-right panel, the four curves show the equations 4 (cyan), 6
  (green) and 7 (magenta) of \citet{ferrarese2002} and the results of
  \citet{baes2003} (red).}
  \label{fig:scale1}
\end{figure*}

\begin{table*}
  \begin{center}
    \label{tab1}
    \begin{tabular}{ccccc}
      \hline
      \hline
      Relation & Normalization ($\alpha$) & Slope ($\beta$) & Scatter & ${\rm Scatter_{corrected}}$ \\
      \hline
      $\log(M_{\rm BH})-M_{\rm K}$            &  -4.37(0.24) &  -0.52(0.01)  &  0.68  &  0.53 \\
      $\log(M_{\rm BH})-M_{\rm B}$            &  -0.61(0.17) &  -0.43(0.01)  &  0.62  &  0.53 \\
      $\log(M_{\rm BH})-\log(\sigma_{\rm c})$ &  -0.26(0.16) &  3.82(0.08)   &  0.42  &  0.28 \\
      $\log(M_{\rm BH})-\log(M_{\rm bulge})$  &  -2.39(0.19) &  0.96(0.02)   &  0.58  &  0.50 \\
      $\log(M_{\rm BH})-\log(V_{\rm c})$      &  -1.61(0.18) &  4.05(0.09)   &  0.45  &       \\
      $\log(M_{\rm BH})-\log(M_{\rm DM})$     &  -8.61(0.42) &  1.35(0.04)   &  0.50  &       \\
      \hline
      \hline
    \end{tabular}
    \caption{Parameters of the linear fits to the scaling relations
      shown in Figure \ref{fig:scale1}.  A correlation of the form
      $y=\alpha+\beta\cdot x$ has been assumed for all relations. The
      uncertainties in the normalizations and in the slopes are shown in
      parentheses. For details about the computation of the Scatter and
      the ${\rm Scatter_{corrected}}$ see Sect. \ref{sub:scale}.  }
    \label{tab:tab1}
  \end{center}
\end{table*}

\begin{table*}
  \begin{center}
    \begin{tabular}{ccccc}
      \hline
      \hline
      Relation & $\alpha$ & $\beta$ & $\gamma$ & Scatter \\
      \hline
      $\log(M_{\rm BH})-M_{\rm K}$            &  17.29(0.10)  &  1.25(0.01)   &  0.04(0.01)  &  0.51 \\
      $\log(M_{\rm BH})-M_{\rm K}$            &  9.81(0.03)   &  0.63(0.01)   &  0.03(0.01)  &  0.47 \\
      $\log(M_{\rm BH})-\log(M_{\rm bulge})$  &  14.16(0.07)  &  -2.21(0.01)  &  0.15(0.01)  &  0.44 \\
      \hline
      \hline
    \end{tabular}
    \caption{Parameters of the fits to the scaling relations shown in
      Figure~\ref{fig:scale2}. A correlation of the form
      $y=\alpha+\beta\cdot x+\gamma\cdot x^2$ has been assumed for all
      three relations. The uncertainties in the parameters are shown in
      parentheses. For details about the computation of the Scatter see
      Sect. \ref{sub:scale}.  }
    \label{tab:tab2}
  \end{center}
\end{table*}

For $f_{\rm Edd}$, which determines the lightcurves
associated with individual quasar events,  we consider instead three
different prescriptions:

\begin {itemize}

\item {\em I}: 
  $f_{\rm Edd}=1$, the simplest possible assumption. Here the quasar is
  either `on' at its maximum Eddington luminosity, or `off'.

\item {\em II}: 
  \begin{equation}
    f_{\rm Edd}(z)=\left\{
    \begin{array}{ll}
      f_{\rm Edd,0}       &  z\geqslant3   \\
      f_{\rm Edd,0}\cdot[(1+z)/4]^{1.4}      &   z<3
    \end{array}
    \right.
  \end{equation}
  with $f_{\rm Edd,0}=0.3$, as suggested by \citet{shankar2004} to match the BH mass function 
  derived from a deconvolution of the AGN luminosity function and the local BH mass function.

\item {\em III}: based on the analysis of self-consistent hydrodynamical
  simulations of galaxy mergers, \citet{hopkins2005} noticed that the
  light curves of active BHs are complex, showing periods of rapid
  accretion after ``first passage'' of the merging galaxies, followed by
  a long-lasting quiescent phase, then a transition to a highly
  luminous, peaked quasar phase, finally a fading away when quasar
  feedback expels gas from the remnant's centre in a self-regulated
  mechanism after the BH reaches a critical mass.  In spite of this
  complexity, as a first order approximation, the typical evolution of
  an active BH can be simply described as a two-stage process of a
  rapid, Eddington-limited growth up to a peak BH mass, preceeded and
  followed by a much longer quiescent phase with lower Eddington ratios.
  In this latter phase, the average time spent by AGN per logarithmic
  luminosity interval can be approximated as \citep{hopkins2005}
    \begin{equation} 
    \frac{{\rm d}t}{{\rm d}\ln L_{\rm bol}}=|\alpha|\,t_9\left(\frac{L_{\rm bol}(t)}{10^9L_\odot}\right)^\alpha,
    \label{eq:dt_hopkins} 
  \end{equation} 
    where $t_9\equiv t_Q(L'>10^9L_\odot)$ and $t_Q(L'>L)$ is the total
  AGN lifetime above a given luminosity $L$; $t_9\sim10^9\,{\rm yr}$
  over the range $10^9L_\odot<L_{\rm bol}<L_{\rm peak}$.  In the range
  $10^{10}L_\odot\lesssim L_{\rm peak}\lesssim 10^{14}L_\odot$,
  \citet{hopkins2005} found that $\alpha$ is a function of only the AGN
  luminosity at the peak of its activity, $L_{\rm peak}$, given by
  $\alpha=-0.95+0.32\log(L_{\rm peak}/10^{12}L_\odot)$, with
  $\alpha=-0.2$ (the approximate slope of the Eddington-limited case) as
  an upper limit. We here interpret the Hopkins model as describing
  primarily the decline phase of the quasar activity, after the black
  hole has grown at the Eddington rate to a peak mass 
  $M_{\rm BH,peak}=M_{\rm BH}(t_{\rm in})+ \mathcal{F} \cdot\Delta M_{\rm BH,Q}\cdot(1-\epsilon)$, 
  where $M_{\rm BH}(t_{\rm in})$ is the initial BH mass and $\Delta M_{\rm BH,Q}$ 
  is the fraction of cold gas mass accreted. Here $\mathcal{F}$ is an additional free parameter, in the range $0 \leq \mathcal{F} \leq 1$.
  For $\mathcal{F} = 1$ the BH emits at the Eddington rate. In the opposite limit ($\mathcal{F}=0$) 
  the AGN reaches instantaneously a peak luminosity, and the whole light curve is  described by equation~(\ref{eq:dt_hopkins}).  
  We found that $\mathcal{F}=0.7$ is the value that best matches the AGN luminosity function. 
  We note that this interpretation of the Hopkins model is plausible but not
  unique, as part of the time described by
  equation~(\ref{eq:dt_hopkins}) could also be associated with the
  rising part of the lightcurve.
  
  From equation (\ref{eq:dt_hopkins}) and with the following definition
  \begin{equation}
    \label{eq:fEdd_tilde}
    \tilde{f}_{\rm Edd}(t) := \frac{L_{\rm bol}(t)}{L_{\rm peak}}
    = f_{\rm Edd}(t)\frac{L_{\rm Edd}(t)}{L_{\rm peak}},  
  \end{equation}
  we can derive:
  \begin{equation}
    \frac{{\rm d}\tilde{f}_{\rm Edd}(t)}{{\rm d}t} = -
    \frac{\tilde{f}_{\rm Edd}^{1-\alpha}(t)}{\alpha t_9}\left(\frac{L_{\rm peak}}{10^9L_\odot}\right)^{-\alpha},
  \end{equation}
  \begin{equation} 
  \Longrightarrow \tilde{f}_{\rm Edd}(t) = \left[\tilde{f}_{\rm Edd,0}^\alpha +
    \left(\frac{L_{\rm peak}}{10^9L_\odot}\right)^{-\alpha}\frac{t}{t_9}\right]^{1/\alpha}.
  \label{eq:fEdd_tilde2}
\end{equation}
  Here we neglected the absolute value of $\alpha$  present in equation~(\ref{eq:dt_hopkins}), 
  for the purpose of having $\tilde{f}_{\rm Edd}(t)$ a decreasing function of time.
  Finally, from equations (\ref{eq:Lagn}), (\ref{eq:fEdd_tilde}) and (\ref{eq:fEdd_tilde2}), we have:
\begin{equation} 
  M_{\rm BH}(t)=M_{\rm BH,peak}+\frac{A}{BC}\left[\left(1+Ct\right)^B-1\right],
  \label{eq:Mbh}
\end{equation}
where $A = \frac{1-\epsilon}{\epsilon}\frac{M_{\rm BH,peak}}{t_{\rm
Edd}}$, $B = \frac{1}{\alpha}+1$, $C = \left(\frac{L_{\rm peak}}{10^9
L_\odot}\right)^{-\alpha}\frac{1}{t_9}$.  To derive equation
(\ref{eq:Mbh}) we set $\tilde{f}_{\rm Edd,0}=1$ for continuity. We also
impose $f_{\rm Edd} =10^{-3}$ as lower limit for the Eddington factor.

\end {itemize}

Figure \ref{fig:accr} shows the evolution of $f_{\rm Edd}(t)$ (top
panel), $M_{\rm BH}(t)$ (central panel) and $L_{\rm bol}(t)$ (bottom
panel) for an illustrative case of a BH of $M_{\rm BH}=10^7M_\odot$
accreting a mass $M_{\rm accr}=5\times 10^8M_\odot$, starting at $z=3$, in the three
prescriptions considered.  The three green curves refer to lightcurve model
{\em III}, in which we set $\mathcal{F}=0.5$ (short dashed), $=0.7$
(dot-long dashed) and $=0.9$ (short dashed-long dashed).

Due to the present uncertainties concerning the origin of the BH seeds
and their mass distribution, we assume $M_{\rm BH,seed}=10^3 M_\odot$
for all seed BHs, irrespective of their halo host properties and their
origin.  Our results are robust with respect to this hypothesis since,
as we have verified, they are basically unaffected by varying $M_{\rm
BH,seed}$ in the range $[10^2-10^5] M_\odot$ at $z\lesssim 3$. More
significant differences occur at higher redshifts, which we will
investigate in detail in future work.

The main parameters of our model are listed in Table 1 of
\citet{croton2006}, with the exception of, as in \citet{delucia2007},
the values for the quiescent hot gas BH accretion rate,
$\kappa_{\rm{AGN}}$ (defined in section \ref{radio}), the star formation
efficiency $\alpha_{\rm{SF}}$ of equation (\ref{m*}), and the
instantaneous recycled fraction of star formation to the cold disk, $R$,
which we set equal to $0.43$ \citep[see Section 3.9 of][]{croton2006}.

\section {Models vs. Observations} \label{sec:comparison}

\subsection {The BH scaling relations} \label{sub:scale}

Several observational evidences indicate that the masses of the BHs
hosted at the centres of galaxies strongly correlate with different
properties of their host bulges and DM haloes. In this section we
compare the most recently observed BH scaling relations at $z=0$ with the
predictions of the original model of \citet{delucia2007}, i.e. the
predictions we obtain when assuming instantaneous mass accretion.  We
explore the effect of specifying the mass accretion rate at the end of
this section.

\subsubsection {One-parameter relations}

In Figure \ref{fig:scale1}, we show the correlation between the masses
of the model BHs with six properties of their hosts, the K- and B-band
bulge magnitude (${\rm M}_{\rm B}$ and ${\rm M}_{\rm K}$), the bulge
mass and velocity dispersion (${\rm M}_{\rm bulge}$ and $\sigma_{\rm
c}$), the circular velocity of the galaxy and the virial mass of the DM
halo (${\rm V}_{\rm c}$ and ${\rm M}_{\rm DM}$).  The blue dots
represent the outputs of the model, while grey and yellow shaded areas
show linear best fits to the model predictions and to the observational
datasets, respectively.

The dots in the plot refer to the population of BHs hosted in the
central galaxies of FOF groups, or subhalos. We do not include those in
satellite galaxies since in this case the host properties cannot be
determined accurately.  The data we have considered are: the
$\rm{M}_{\rm BH}-{\rm M}_{\rm B}$ and ${\rm M}_{\rm BH}-{\rm M}_{\rm K}$
relations of \citet{marconi2004} (top panels) the ${\rm M}_{\rm
BH}-\sigma_{\rm c}$ relation of \citet{ferrarese2005} (central left) the
${\rm M}_{\rm BH}-{\rm M}_{\rm bulge}$ relation of \citet{haring2004}
(central right) and the ${\rm M}_{\rm BH}-{\rm V}_{\rm c}$ relation of
\citet{baes2003} (bottom left).  No direct observational estimate is
available for the ${\rm M}_{\rm BH}-{\rm M}_{\rm DM}$ relation shown in
the bottom right panel. The curves shown in this panel have been derived
using different assumptions for the ${\rm M}_{\rm DM}-{\rm V}_{\rm c}$
relation. In particular, the cyan, green and magenta lines correspond to
equations (4), (6) and (7) of \citet{ferrarese2002}, while the red curve
is taken from \citet{baes2003}.

Model predictions for ${\rm V}_{\rm c}$ and $\sigma_{\rm c}$ have been
obtained by adopting two different assumptions: i) ${\rm V}_{\rm c}={\rm
V}_{\rm max}$, where ${\rm V}_{\rm max}$ is the maximum rotational
velocity of the subhalo hosting the galaxy at its centre, and ii) ${\rm
V}_{\rm c}=1.8\, {\rm V}_{\rm vir}$ as derived by \citet{seljak2002}.
The bulge velocity dispersion $\sigma_{\rm c}$ is derived from the ${\rm
V}_{\rm c}-\sigma_{\rm c}$ relation of \citet{baes2003}. In the bottom
panels, the grey areas correspond to a circular velocity obtained
through hypothesis i) while the green ones, in better agreement with the
data, assume hypothesis ii).

The linear fit to the model data has been obtained using the bisector
modification to the ordinary least squares minimization approach,
proposed by \cite{akritas1996}, for which the best-fit results
correspond to the bisection of those obtained from minimizations in the
vertical and horizontal directions. The estimator is robust and has the
advantage of taking into account the possible intrinsic scatter in the
relation.  The values of the best fit slope and the normalization are
listed in Table \ref{tab:tab1} along with the scatter around the best
fitting line.  The uncertainties of the best fit parameters, also
reported in the table, have been obtained by imposing $\chi^2_{\rm
d.o.f.}=1$.

As can be seen in Figure \ref{fig:scale1}, the best fits to the model
agree well with that to the data, within the scatter.  We note that, in
all relations plotted, the scatter in the model is larger than that of
the real data and also larger than the internal scatter observed in
similar relations obtained from the recent hydrodynamical simulations of
galaxy mergers \citep[see e.g.][]{hopkins2007a}. 
However, we notice that a large fraction of our model BHs are found in low-mass systems
for which the scatter in the scaling relation is large.
On the contrary, in the real datasets (and hydro-simulations) the majority
of BHs belong to massive galaxies for which, according to our model,
the scatter in the scaling relation is significantly smaller.
To investigate whether the difference in the intrinsic scatter is real or is induced by a different
sampling of the BH population, for each BH scaling relation we have discretized 
the range of the observed host galaxy properties in finite bins and generated 500 sub-samples by
randomly extracting $N_{\rm obs}(\Delta_X)$ model BHs from the parent catalogue,
where $N_{\rm obs}(\Delta_X)$ is the number of BHs in the real dataset in each bin $\Delta_X$. 
We have repeated the same fitting procedure in the 500 sub-samples and found
that the scatter is significantly reduced in this exercise, as indicated
in the last column of Table \ref{tab:tab1}, that lists the average
scatter in the sub-catalogues. 
Therefore, the mismatch in the scatter results from sampling different BH populations: 
small objects in the model, massive objects in the observations.
Moreover, for the ${\rm M}_{\rm BH}-\sigma_{\rm c}$
relation the scatter is very close to 0.21, which is the value measured
by \cite{hopkins2007a} both in the observed and simulated data.

\subsubsection {Non-linear fits}

\begin{figure}
  \includegraphics[width=0.45\textwidth]{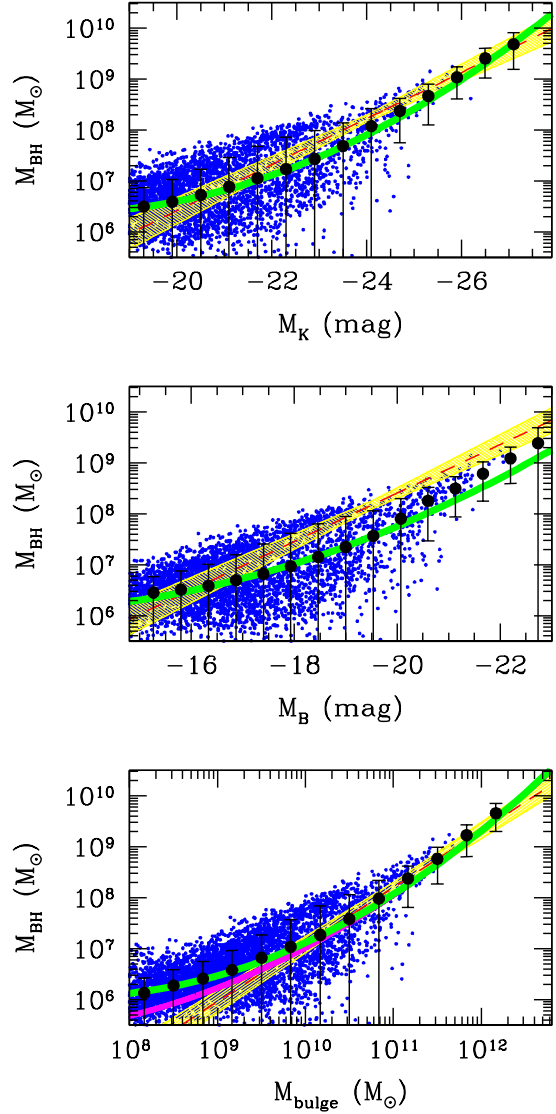}
  \caption{The tree model scaling relations best constrained by observations.
    Here the black dots (with error bars) represent the medians (and the corresponding
    first and third quartiles) of the model outputs, computed in a
    discrete number of bins.  The green lines show the best
    three-parameters fits to the model outputs (blue points). The
    magenta line in the lower panel refers to the best-fit relation
    obtained by \citet{wyithe2006}.  }
  \label{fig:scale2}
\end{figure}

The agreement between model and data is satisfactory. However, we need
to keep in mind that the model predictions for $V_{\rm c}$ and
$\sigma_{\rm c}$ and the observed relation between $\log(M_{\rm BH})$
and $\log(M_{\rm DM})$ have been obtained assuming further theoretical
hypotheses.  Consequently, the more constraining and reliable relations
are the ones between the BH masses and the bulge magnitudes and
masses. Focusing on these relations and thanks to the huge number of
model BHs, we have been able to investigate whether a non-linear fit
provides a better match to the data. We find that the best fit is a
quadratic function, $y=\alpha+\beta\cdot x+\gamma \cdot x^2$. Figure
\ref{fig:scale2} shows this fit (heavy green lines), together with the
medians, the first and third quartiles (black points with error bars) of
the model output, computed in a discrete number of bins.  The internal
scatter is significantly smaller than in the linear fit case.  The
values of the best fit parameters are reported in Table \ref{tab:tab2}.
While we predict, on average, too low BH masses for a fixed $M_{\rm B}$
with respect to the observations (still consistent within the errors)
the model predictions are in very good agreement with the data for the
$\log(M_{\rm BH})-M_{\rm K}$ and $\log(M_{\rm BH})-\log(M_{\rm bulge})$
relations. Interestingly, the 3-parameters fit of the latter relation is
in excellent agreement with the one found by \citet{wyithe2006} (magenta
solid line in lower panel of Figure \ref{fig:scale2}).

\begin{figure*}
  \includegraphics[width=1.0\textwidth]{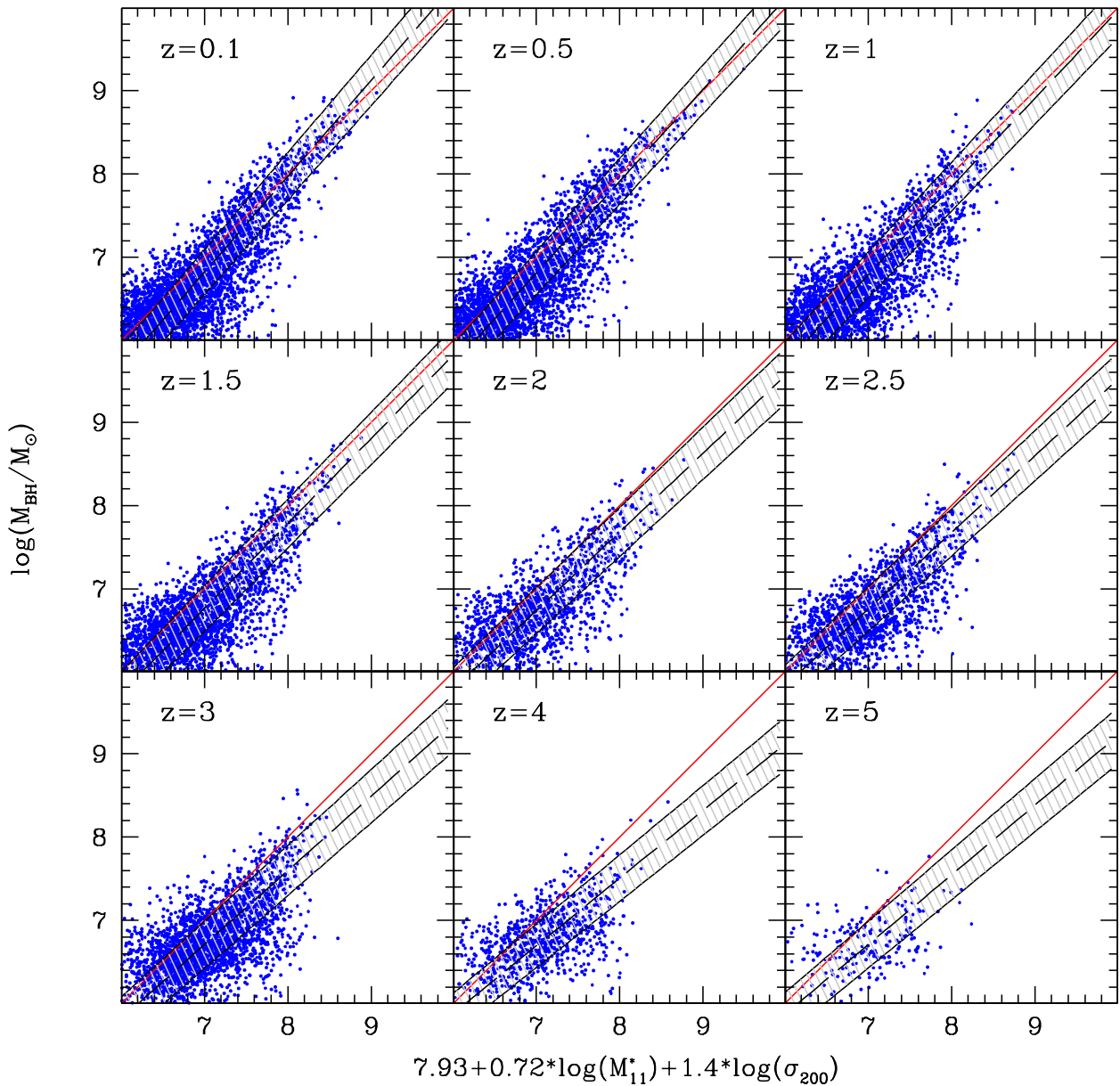}
  \caption{The BH fundamental plane in the redshift range $0.1\leq
    z\leq5$. The blue dots are the model outputs, while the grey shaded
    areas show the best-fits to them. The red lines, corresponding to
    the bisectors of the plots, are the predictions of
    \citet{hopkins2007a}. The galaxy stellar mass, $M_{11}^*$, is given
    in units of $10^{11}M_\odot$, while the bulge velocity dispersion,
    $\sigma_{200}$, is in units of 200 km $s^{-1}$.}
  \label{fig:bhfp}
\end{figure*}

\subsubsection {The fundamental plane relation}

In Figure \ref{fig:bhfp} we compare the BH fundamental plane relation of
our model at different redshifts with that obtained by
\citet{hopkins2007a} using both observational data and the outputs of
hydrodynamical simulations of galaxy mergers:
\begin{equation*}
\log(M_{\rm BH}/M_\odot) = 7.93+0.72\log(M_{11}^*)+1.4\log(\sigma_{200}),
\end{equation*}
where $M_{11}^*$ is the galaxy stellar mass in units of
$10^{11}M_\odot$, and $\sigma_{200}$ is the bulge velocity dispersion in
units of 200 km $s^{-1}$.  The red lines, bisectors of the plots, show
the fundamental plane relation proposed by \citet{hopkins2007a}. Model
prediction are represented by blue dots, the black line is the best fit
to the model and the shaded area its $1 \sigma$ scatter.  At low
redshifts the agreement is very good.  This is not surprising since at
$z\sim 0$ our model agrees with the ${\rm M}_{\rm BH}-{\rm M}_{\rm
bulge}$ and ${\rm M}_{\rm BH}-\sigma_{\rm c}$ scaling relations that
represent fundamental plane projections. A discrepancy appears
at high redshifts. However, at $z>3$ the fit involves only few objects
and therefore may not be very significant, especially when we account
for the non-zero intrinsic scatter in the fundamental plane proposed by
\citet{hopkins2007a}.  
A remarkable success of our model is that it
predicts very little evolution of the fundamental plane relation, at
least out to $z=3$, in agreement with \citet{hopkins2007a}.  The
intrinsic scatter, which does not evolve with time either, is 3 times
larger than in \citet{hopkins2007a} (we found a value around $0.6$ at
all redshifts, while the one reported by \citet{hopkins2007a} is about
$0.2$).  As discussed previously, the mismatch is reduced when using a
number of model BHs consistent with the observed one.

\subsubsection {Dependence on the accretion history} \label{sub:dephist}

\begin{figure}
  \includegraphics[width=0.5\textwidth]{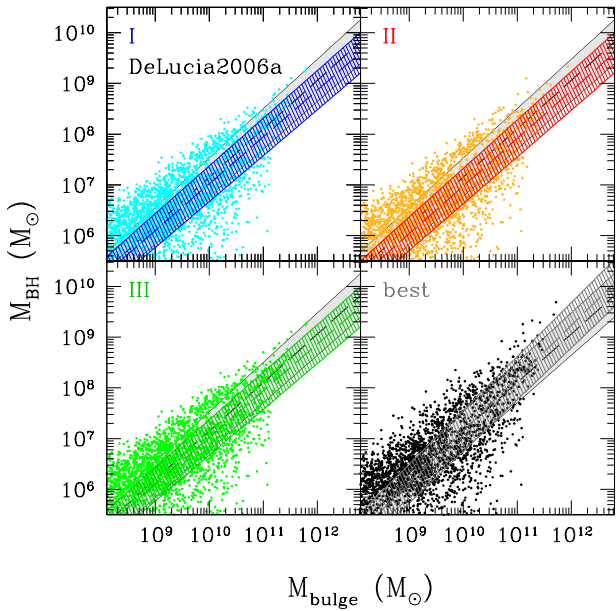}
  \caption{The $\log(M_{\rm BH})-\log(M_{\rm bulge})$ scaling relation
    for our different prescriptions for the BH mass accretion. The
    filled dots represent model predictions, the grey shaded areas show
    the linear fit to the {\em DeLucia2006a} model scaling relation and
    the other hatched areas indicate the linear fit to the {\em I}, {\em
    II} and {\em III} lightcurve models, as indicated by the labels.
    The black dots and grey shaded areas, in the lower right panel, show
    the prediction obtained with the parameterization given by the
    equations (\ref{eq:best}), as explained in Section \ref{sec:AGN_LF}.
    }
\label{fig:scale1.1}
\end{figure}

All scaling relations predicted by our model assume that BHs accrete
mass instantaneously after merging events. What happens if we relax this
assumption and specify the mass accretion rate instead?
Figure~\ref{fig:scale1.1} shows the impact of adopting different
accretion recipes on the ${\rm M}_{\rm BH}-{\rm M}_{\rm bulge}$
relation.  As usual, filled dots represent model predictions, grey
shaded areas show the linear fit to the {\em DeLucia2006a} model scaling
relation and the other hatched areas indicate the linear fit to the
model predictions obtained with our different recipes, as indicated by
the labels\footnote{The meaning of the black dots and shaded areas in the bottom left panel of
Fig. 5 is discussed in Section \ref{sec:AGN_LF}.}.
Clearly, these predictions depend little on the assumed mass accretion histories
for each individual quasar event (the fit parameters have fluctuations
of no more than about $1\%$).  This is a consequence of the fact that
the BH scaling relations depend mainly on the total mass accreted, and
very little on the time spent in the accretion process. We have verified
that all other scaling relations, including also the fundamental plane
relation, does not change significantly by adopting any of the mass accretion
prescriptions described in Section \ref{sec:new}.

\subsection {The BH mass function}

\begin{figure}
  \includegraphics[width=0.45\textwidth]{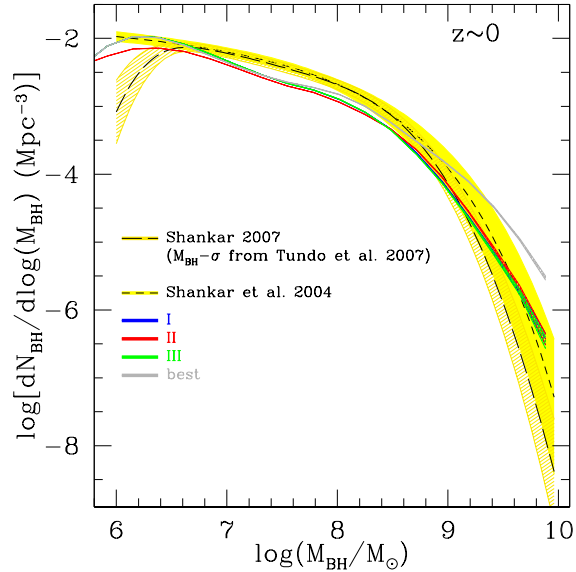}
  \caption{ Comparison of the BH mass function predicted by lightcurve
    models {\em I}, {\em II} and {\em III} with the one observationally derived by
    \citet{shankar2004}, and with the new one obtained by Shankar
    (2007, in preparation) using the $M_{\rm BH}-\sigma$ relation by
    \citet{tundo2007}.  The grey areas show the prediction obtained
    with the parameterization given by the equations (\ref{eq:best}), as
    explained in Section \ref{sec:AGN_LF}.  }
  \label{fig:bhmf}
\end{figure}

The BH mass function (MF) is defined as the differential co-moving
number density of BHs as a function of their mass.  In
Figure~\ref{fig:bhmf}, we compare the BH MF predicted by our model for the
prescriptions {\em I} (blue line), {\em II} (red) and {\em III} (green)
with those observed by \citet{shankar2004} (grey area) and by Shankar
(2007, in preparation) (yellow area) at $z\sim 0$.  In neither case the
BH masses were determined directly: \citet{shankar2004} derive the BH
mass from the observed ${\rm M}_{\rm BH}-L_{\rm bulge}$ relation while
Shankar (2007) use the ${\rm M}_{\rm BH}-\sigma_{\rm c}$ relation
of \citet{tundo2007}.  We note that the model BH MF is in good agreement
with the observed ones, within the mass range accessible to
observations exept in the interval $\sim10^7-10^9 M_\odot$, in which the number
density of model BHs is smaller than the observed one.

The reason of the small mismatch between the observed and the model BH MFs
will be investigated in a forthcoming paper in which we study the
redshift evolution of the BH MF and its dependency on the properties of the
host galaxy. Finally, we note that, as shown in Figure~\ref{fig:bhmf},
the model predictions for the BH MF are robust with respect
to the prescription adopted for the mass accretion history
of the individual quasar episodes.

\subsection {The AGN bolometric luminosity function} \label{sec:AGN_LF}

\begin{figure*}
\includegraphics[width=1.0\textwidth]{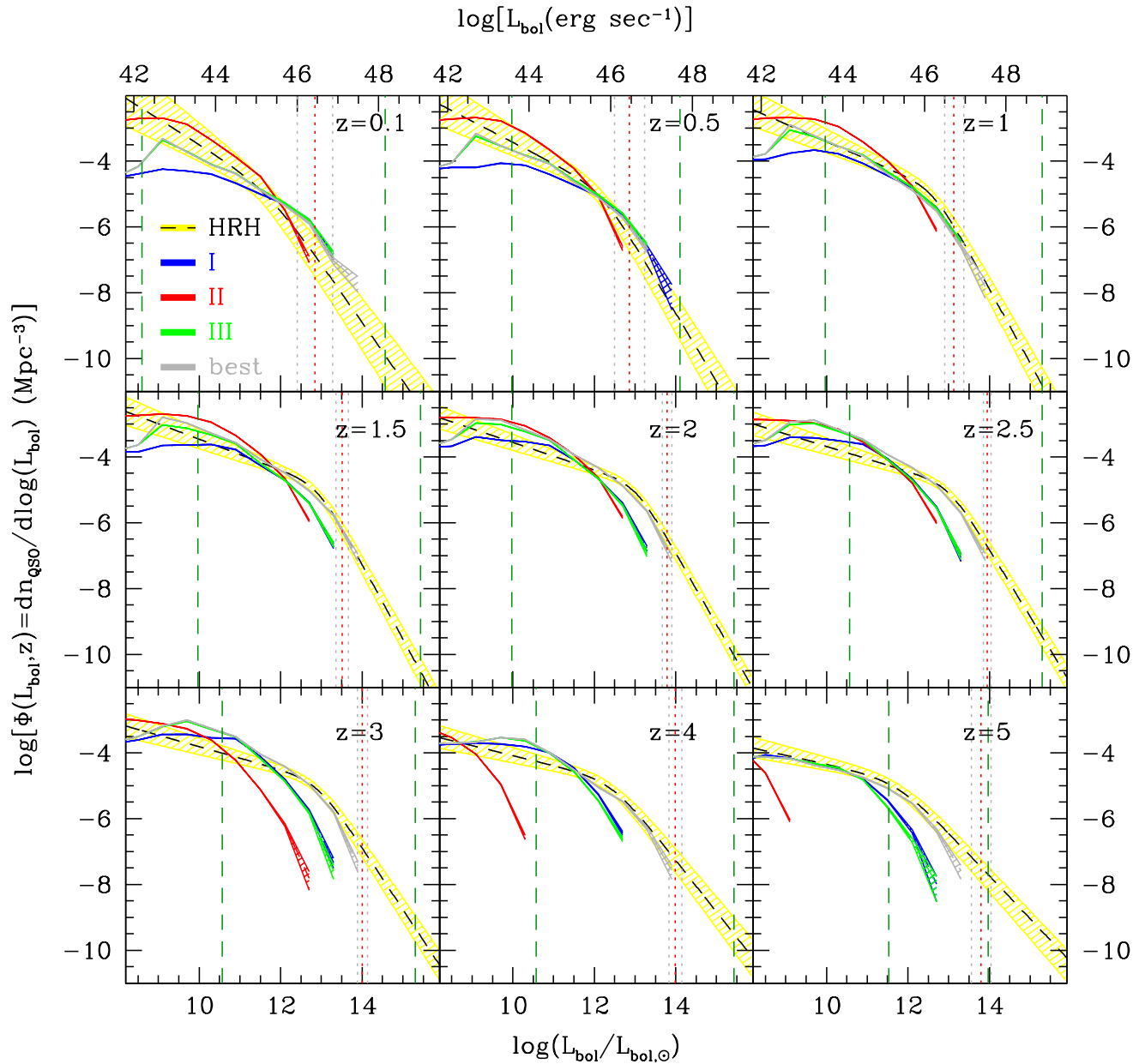}
\caption{The bolometric LFs predicted by our lightcurve models {\em I}
  (blue bands), {\em II} (red bands) and {\em III} (green bands), in the
  redshift range $0.1\leq z\leq5$, are here compared with the best-fits
  to observational data obtained by \citet{hopkins2007} (yellow
  bands). The grey areas show the predictions obtained with the
  parameterization given by the equations (\ref{eq:best}), as explained
  in Section \ref{sec:AGN_LF}.  Uncertainties in the model LFs are
  computed by assuming Poisson statistics.  The dashed vertical green
  lines mark the range of the bolometric luminosities accessible to
  observations. The dotted red vertical lines show the luminosities
  beyond which the LF of \citet{hopkins2007} predicts a number of AGN in
  the whole volume of our simulation smaller than 10. The vertical grey
  dotted lines around the red ones have been calculated considering the
  error in the best-fit of \citet{hopkins2007}.  }
\label{fig:lf}
\end{figure*}

The luminosity function (LF) of AGN, namely the derivative of their
co-moving number density with respect to luminosity, represents a unique
tool to understand their cosmological evolution.  Semi-analytic models
predict the total (bolometric) luminosity of a statistically complete
AGN catalogue, and to compare model LFs with observations we need to
specify a bolometric correction, i.e. how to convert the luminosities
observed in a particular band into bolometric ones
\citep{elvis1994,marconi2004,hopkins2006}.  Another correction is
required to account for possible incompleteness effects \citep[see
e.g.][]{comastri2004,gilli2007}, which includes the possible existence
of a population of obscured AGN whose fraction may depend on the
wavelength band and redshift
\citep{elvis1994,marconi2004,lafranca2005,lamastra2006}.

Here we compare our predictions with the bolometric LF obtained by
\citet{hopkins2007} from the LFs observed in different bands: radio
\citep[see e.g.][]{nagar2005}, optical \citep[see e.g.][]
{kennefick1995,schmidt1995,kohler1997,grazian2000,fan2001a,wolf2003,hunt2004,cristiani2004,
croom2005,richards2005,richards2006,siana2006,fontanot2007,shankar2007,bongiorno2007},
infra-red \citep[see e.g.][]{brown2006,matute2006,babbedge2006}, soft
X-ray \citep[see
e.g.][]{miyaji2000,miyaji2001,silverman2005b,hasinger2005}, hard X-ray
\citep [see
e.g.][]{barger2003a,ueda2003,barger2003b,nandra2005,sazonov2004,
silverman2005a,lafranca2005,shinozaki2006,beckmann2006}, and from
emission lines \citep[see e.g.][]{hao2005}.  Uncertainties in these
corrections contribute to the scatter in the observed LF, i.e. to the
width of the yellow areas in Figure \ref{fig:lf} that show the AGN
bolometric LF of \citet{hopkins2007} at different redshifts. The model
predictions are also represented by areas with different colours, with a
width corresponding to $1\sigma$ Poisson error bars.  The vertical,
green dashed lines bracket the bolometric luminosity range accessible to
observations. The vertical, red dotted lines show the luminosities
beyond which the LF of \citet{hopkins2007} predicts less than 10 AGN in
the volume of our simulation, i.e. the maximum luminosities at which our
model BH sample is statistically meaningful; $1\sigma$ uncertainties on
this maximum luminosity are represented by the two grey dotted lines.

From Figure \ref{fig:lf} we see that, on average, type-{\em I} lightcurve
underestimates the AGN number density at all epochs.  However,
while at high redshifts the model matches the faint-end of the LF and
underpredicts the number density of the bright objects, the situation is
completely reversed at $z\sim 0$, where the model correctly reproduces
the number density of bright AGN but underestimates the faint ones.  At
low redshifts the problem can be alleviated by reducing the Eddington
factor, as in our type-{\em II} lightcurve. However, in this case the
discrepancy between model and data at high redshifts increases.
Adopting the type-{\em III} lightcurve allows to match observations
in the whole range of luminosities in the redshift range
$0.5\lesssim z\lesssim 1$, but overestimates the number of luminous AGN
at $z\lesssim 0.5$ and underestimates them at $z\gtrsim1$.

Therefore, we conclude that in our present semi-analytical framework we
can reproduce the observed AGN LF at low and intermediate redshifts.
However, at $z\gtrsim 1$, we under-predict the number density of bright
AGN, regardless of the BH mass accretion rate and light curve model
assumed for each quasar episode.
To investigate if it is possible to modify our prescription for the mass accretion to fit the AGN LF at
all redshifts, we tried different values of $f_{\rm Edd}$ and $\epsilon$ as a function of $t$ and $M_{\rm BH}$, 
within physically motivated ranges.
Despite of the considerable freedom in choosing $f_{\rm
Edd}(t,M_{\rm BH})$ we failed to find a model able to match
simultaneously the observed BH scaling relations, the BH MF, and the AGN
LF, especially at high redshifts.  We also used different 
plausible values for the BH seed mass, and we still were not able to
fit the high-z LF.  We interpret this failure as an
indication that our theoretical framework itself is inadequate to
account fully successfully for the AGN phenomenon.

One possible way out is to modify the model assumptions for the
efficiency of BH growth in the {\em quasar mode} following mergers at high z.
A significant improvement of our results at high redshifts can
for example be obtained by substituting equation (\ref{eq:accretionQ})
and (\ref{eq:fBH}) with
\begin{equation}
  \left\{
    \begin{array}{ll}
      f_{\rm BH}=0.01\cdot \log\left(\frac{M_{\rm BH}}{10^3 M_\odot}+1\right)\cdot z  
      &  z>1.5 \,{\rm and}\, M_{\rm BH}>10^6M_\odot \\
      {\Delta}M_{\rm BH,Q} = 0.01\cdot m_{\rm cold}  &  z>6
    \end{array}
    \right.
    \label{eq:best}
\end{equation}
while keeping prescription {\em III} for the quasar lightcurves.  
The predictions of this new model for the $\log(M_{\rm BH})-\log(M_{\rm bulge})$ scaling relation is shown
as black dots in the  bottom-right plot of Fig. \ref{fig:scale1.1}.
Model predictions for the BH MF and AGN LF are shown in Figures \ref{fig:bhmf} and \ref{fig:lf}, respectively.
An accretion efficiency that increases with the redshift
has been already advocated in the {\em dynamical model}
of \citet{croton2006b}.
A physical justification to this assumption is provided
by \citet{mo1998}. Indeed, their model predicts that galactic disks were
more centrally concentrated in the past, making
it more efficient the BH feeding at high redshift.
It is worth stressing that equation (\ref{eq:best}) might not provide the best fit to the data as we did
not explore the parameter space systematically.  However, it suggests
that a good match to the observed scaling relations, BH MF and AGN LFs
can be obtained within our semi-analytic framework by modest changes of
the BH growth at high redshifts.  The solution provided by
equation (\ref{eq:best}) is not unique either, since larger amounts of
mass can be accreted also by invoking alternative mechanisms that
trigger gas accretion episodes, for example by secular evolution through
disk instabilities, or by alluding to a higher gas cooling efficiency
\citep[see e.g.][]{viola2007}.

\section {Conclusions} \label{sec:conclusions}
  
In this paper we have used and extended a semi-analytic model for the
co-evolution of galaxies and their central BHs, developed on the outputs
of the Millennium Simulation \citep{springel2005}, and described in
detail in \citet{croton2006} and \citet{delucia2007}. The aim of the model is to reproduce  
the observed properties of BHs, AGN and their galaxy
hosts.  The physical assumptions in the model with respect to BH growth
can be divided into two sets. The first one concerns the mass accretion
history of the central BHs in halos, where we distinguish between {\em
radio mode} and {\em quasar mode} \citep{croton2006}.  This set makes
predictions for the relation between BH and galaxy host properties,
which can be compared to the observed scaling relations between BH mass
and different properties of their host galaxies. The second set of
prescriptions specifies the detailed AGN activity and lightcurve of
individual quasar episodes, and leads to predictions for the AGN LF as a
function of redshift. We considered three different models for this
detailed AGN activity, one of them motivated by the results of recent
hydrodynamical simulations of galaxy mergers that include BH growth and
feedback \citep{hopkins2005,dimatteo2005,springel2005c}.

\noindent The main results  of our analysis are as follows:

{\em (i)} The semi-analytic model is approximately able to reproduce the
observed BH scaling relations over the whole range of BH masses and
galaxy properties probed by observations.  The intrinsic scatter in the
model is significantly larger than in the data, a mismatch that can in
part be accounted for by adopting the observational selection criteria
to obtain a mock BH catalogue with similar characteristics as the
observed one.

{\em (ii)} We find evidence that a quadratic relationship provides a
significantly better fit to some of the model scaling relationships than
a linear one, as already noticed by \citet{wyithe2006}.

{\em (iii)} Our model also matches the BH fundamental plane
relation derived by \citet{hopkins2007a}, and successfully predicts very
little evolution of this plane, at least out to $z\sim 3$.

{\em (iv)} The model BH mass function is in good agreement with the
observed one within the mass range accessible by observations, 
except on the range $\sim10^7-10^9 M_\odot$, in which the number density
predicted by the model is smaller than the observed one.

{\em (v)} Model predictions for the BH mass function, scaling relations
and fundamental plane relation are basically unaffected when using
different prescriptions for the AGN lightcurves of individual quasar
events. This is because these predictions are only sensitive to the
model assumptions for the absolute growth of the BHs in each
merger event.

{\em (vi)} The AGN LF is systematically underestimated by assuming that
BHs accrete mass with a constant Eddington factor $f_{\rm Edd}=1$. 
The detail of the discrepancy, however, change with redshift since at high $z$
the model matches the faint-end of the LF but underpredicts the number
density of the brightest objects, while the situation is reversed at
$z\sim 0$, in agreement with the results of several semi-analytic models
\citep[see, e.g.][and references therein]{marulli2007}.  
Reducing the Eddington ratio, as in our lightcurve model {\em II}, alleviates
the faint-end mismatch but amplifies the
bright-end discrepancy at high redshifts.  A significant improvement at
low redshifts is obtained when the Eddington-limited growth of the BH 
is followed by a long quiescent phase with lower Eddington ratios, 
as suggested by \citet{hopkins2005} and implemented in our
lightcurve model {\em III}.
In this case our model is able to match the
observed AGN LF in the interval $ 0.1\lesssim z\lesssim 1$, over the
whole range of luminosities that are accessible to observations and
where our predictions are statistically significant.  However, our
predicted number density of bright AGN is still biased low at
$z\gtrsim1$.

{\em (vii)} Our model is able to account for all observations considered
in this work apart for the AGN LF at high redshifts.  We were not able
to eliminate this mismatch by simply modifying the accretion efficiency, $\epsilon$,  
the Eddington factor, $f_{\rm Edd}$, or the BH seed mass (when considered in physically plausible 
ranges).
Clearly, we need to modify assumptions in the underlying semi-analytic framework for BH
growth. A simple, {\em ad hoc} increase of the mass fraction accreted
during the {\em quasar mode} at high redshift can indeed remedy the
problem. However, this solution is not unique as several high-redshift
modifications to the original model, like new mechanisms that trigger BH
activity in addition to galaxy merging or more efficient gas cooling
resulting in a larger reservoir of cold gas, can be advocated to bring
the predictions in line with observations.  However, it remains to be
seen whether any of these alternatives is physically plausible.

{\em (viii)} Our model predictions at $z<3$ are robust to changes in the
assumed BH seed mass, but are sensitive to it at
larger redshift. We will further explore this issue in a subsequent
paper where we plan to study to what extent current observations can
constrain the seed BH mass function.

From our analysis we conclude that the AGN LF at high redshifts
constitutes a strong constraint for semi-analytic models that describe
the co-evolution of galaxies, BHs and AGN.  This suggests that
significant improvements can be obtained in two ways.  From the
theoretical side, we need to develop a physically motivated mechanism
capable of increasing the number density of bright AGN at $z\gtrsim1$
without modifying the model predictions at low redshifts.  From the
observational point of view, we need to improve the AGN LF estimates at
high redshift, both by enlarging current high-$z$ AGN samples and by
reducing the current uncertainties originating from bolometric and
incompleteness corrections, in particular for the population of Compton
Thick AGN.  In addition, other observational tests should be performed,
like the ability of our model to match the observed AGN clustering, as
quantified by the angular and spatial two-point correlations function.
In particular, \citet{lidz2006} pointed out that the luminosity
dependence of quasar clustering can discriminate between different
lightcurve models, a question we will address in a forthcoming work.

\section*{acknowledgments}
We thank Simon D. M. White, Gabriella De Lucia, Andrea Merloni, Philip Hopkins, Francesco Shankar and
Umberto Maio for very useful discussions.  FM thanks the
Max-Planck-Institut f\"ur Astrophysik for the kind hospitality.  We
acknowledge financial contribution from contracts ASI-INAF I/023/05/0,
ASI-INAF I/088/06/0 and INFN PD51.  SB acknowledges the PhD fellowship
 of the International Max Planck Research 
School in Astrophysics, and the support received from a Marie Curie Host 
Fellowship for Early Stage Research Training

\bibliographystyle{mn2e}
\bibliography{bib}

\begin{thebibliography}{}

\bibitem[\protect\citeauthoryear{{Akritas} \& {Bershady}}{{Akritas} \&
  {Bershady}}{1996}]{akritas1996}
{Akritas} M.~G.,  {Bershady} M.~A.,  1996, \apj, 470, 706

\bibitem[\protect\citeauthoryear{{Babbedge} et~al.,}{{Babbedge}
  et~al.}{2006}]{babbedge2006}
{Babbedge} T.~S.~R.,  et~al., 2006, \mnras, 370, 1159

\bibitem[\protect\citeauthoryear{{Baes}, {Buyle}, {Hau} \& {Dejonghe}}{{Baes}
  et~al.}{2003}]{baes2003}
{Baes} M.,  {Buyle} P.,  {Hau} G.~K.~T.,    {Dejonghe} H.,  2003, \mnras, 341,
  L44

\bibitem[\protect\citeauthoryear{{Barger}, {Cowie}, {Capak}, {Alexander},
  {Bauer}, {Brandt}, {Garmire} \& {Hornschemeier}}{{Barger}
  et~al.}{2003}]{barger2003b}
{Barger} A.~J.,  {Cowie} L.~L.,  {Capak} P.,  {Alexander} D.~M.,  {Bauer}
  F.~E.,  {Brandt} W.~N.,  {Garmire} G.~P.,    {Hornschemeier} A.~E.,  2003,
  \apjl, 584, L61

\bibitem[\protect\citeauthoryear{{Barger} et~al.,}{{Barger}
  et~al.}{2003}]{barger2003a}
{Barger} A.~J.,  et~al., 2003, \aj, 126, 632

\bibitem[\protect\citeauthoryear{{Beckmann}, {Soldi}, {Shrader}, {Gehrels} \&
  {Produit}}{{Beckmann} et~al.}{2006}]{beckmann2006}
{Beckmann} V.,  {Soldi} S.,  {Shrader} C.~R.,  {Gehrels} N.,    {Produit} N.,
  2006, \apj, 652, 126

\bibitem[\protect\citeauthoryear{{Benson}, {Bower}, {Frenk}, {Lacey}, {Baugh}
  \& {Cole}}{{Benson} et~al.}{2003}]{benson2003}
{Benson} A.~J.,  {Bower} R.~G.,  {Frenk} C.~S.,  {Lacey} C.~G.,  {Baugh} C.~M.,
     {Cole} S.,  2003, \apj, 599, 38

\bibitem[\protect\citeauthoryear{{Binney} \& {Tremaine}}{{Binney} \&
  {Tremaine}}{1987}]{binney1987}
{Binney} J.,  {Tremaine} S.,  1987, {Galactic dynamics}.
Princeton, NJ, Princeton University Press, 1987, 747 p.

\bibitem[\protect\citeauthoryear{{Bongiorno} et~al.,}{{Bongiorno}
  et~al.}{2007}]{bongiorno2007}
{Bongiorno} A.,  et~al., 2007, preprint, astro-ph/0704.1660, 704

\bibitem[\protect\citeauthoryear{{Brown} et~al.,}{{Brown}
  et~al.}{2006}]{brown2006}
{Brown} M.~J.~I.,  et~al., 2006, \apj, 638, 88

\bibitem[\protect\citeauthoryear{{Cattaneo}, {Blaizot}, {Devriendt} \&
  {Guiderdoni}}{{Cattaneo} et~al.}{2005}]{cattaneo2005}
{Cattaneo} A.,  {Blaizot} J.,  {Devriendt} J.,    {Guiderdoni} B.,  2005,
  \mnras, 364, 407

\bibitem[\protect\citeauthoryear{{Cavaliere} \& {Vittorini}}{{Cavaliere} \&
  {Vittorini}}{2002}]{cavaliere2002}
{Cavaliere} A.,  {Vittorini} V.,  2002, \apj, 570, 114

\bibitem[\protect\citeauthoryear{{Cimatti}, {Daddi} \& {Renzini}}{{Cimatti}
  et~al.}{2006}]{cimatti2006}
{Cimatti} A.,  {Daddi} E.,    {Renzini} A.,  2006, \aap, 453, L29

\bibitem[\protect\citeauthoryear{{Ciotti} \& {Ostriker}}{{Ciotti} \&
  {Ostriker}}{2007}]{ciotti2007}
{Ciotti} L.,  {Ostriker} J.~P.,  2007, \apj, 665, 1038

\bibitem[\protect\citeauthoryear{{Colless} et~al.,}{{Colless}
  et~al.}{2001}]{colless2001}
{Colless} M.,  et~al., 2001, \mnras, 328, 1039

\bibitem[\protect\citeauthoryear{{Comastri}}{{Comastri}}{2004}]{comastri2004}
{Comastri} A.,  2004, in {Barger} A.~J.,  ed., Astrophysics and Space Science
  Library Vol.~308 of Astrophysics and Space Science Library, {Compton-Thick
  AGN: The Dark Side of the X-Ray Background}.
pp 245--+

\bibitem[\protect\citeauthoryear{{Cowie} \& {Binney}}{{Cowie} \&
  {Binney}}{1977}]{cowie1977}
{Cowie} L.~L.,  {Binney} J.,  1977, \apj, 215, 723

\bibitem[\protect\citeauthoryear{{Cowie}, {Songaila}, {Hu} \& {Cohen}}{{Cowie}
  et~al.}{1996}]{cowie1996}
{Cowie} L.~L.,  {Songaila} A.,  {Hu} E.~M.,    {Cohen} J.~G.,  1996, \aj, 112,
  839

\bibitem[\protect\citeauthoryear{{Cox}}{{Cox}}{2004}]{cox2004}
{Cox} T.~J.,  2004, Ph.D.~Thesis, University of California, Santa Cruz

\bibitem[\protect\citeauthoryear{{Cristiani} et~al.,}{{Cristiani}
  et~al.}{2004}]{cristiani2004}
{Cristiani} S.,  et~al., 2004, \apjl, 600, L119

\bibitem[\protect\citeauthoryear{{Croom} et~al.,}{{Croom}
  et~al.}{2005}]{croom2005}
{Croom} S.~M.,  et~al., 2005, \mnras, 356, 415

\bibitem[\protect\citeauthoryear{{Croton}}{{Croton}}{2006}]{croton2006b}
{Croton} D.~J.,  2006, \mnras, 369, 1808

\bibitem[\protect\citeauthoryear{{Croton} et~al.,}{{Croton}
  et~al.}{2006}]{croton2006}
{Croton} D.~J.,  et~al., 2006, \mnras, 365, 11

\bibitem[\protect\citeauthoryear{{De Lucia} \& {Blaizot}}{{De Lucia} \&
  {Blaizot}}{2007}]{delucia2007}
{De Lucia} G.,  {Blaizot} J.,  2007, \mnras, 375, 2

\bibitem[\protect\citeauthoryear{{De Lucia}, {Kauffmann} \& {White}}{{De Lucia}
  et~al.}{2004}]{delucia2004}
{De Lucia} G.,  {Kauffmann} G.,    {White} S.~D.~M.,  2004, \mnras, 349, 1101

\bibitem[\protect\citeauthoryear{{De Lucia}, {Springel}, {White}, {Croton} \&
  {Kauffmann}}{{De Lucia} et~al.}{2006}]{delucia2006}
{De Lucia} G.,  {Springel} V.,  {White} S.~D.~M.,  {Croton} D.,    {Kauffmann}
  G.,  2006, \mnras, 366, 499

\bibitem[\protect\citeauthoryear{{Di Matteo}, {Colberg}, {Springel},
  {Hernquist} \& {Sijacki}}{{Di Matteo} et~al.}{2007}]{dimatteo2007}
{Di Matteo} T.,  {Colberg} J.,  {Springel} V.,  {Hernquist} L.,    {Sijacki}
  D.,  2007, ArXiv e-prints, 705

\bibitem[\protect\citeauthoryear{{Di Matteo}, {Springel} \& {Hernquist}}{{Di
  Matteo} et~al.}{2005}]{dimatteo2005}
{Di Matteo} T.,  {Springel} V.,    {Hernquist} L.,  2005, \nat, 433, 604

\bibitem[\protect\citeauthoryear{{Efstathiou}}{{Efstathiou}}{1992}]{efstathiou%
1992}
{Efstathiou} G.,  1992, \mnras, 256, 43P

\bibitem[\protect\citeauthoryear{{Efstathiou} \& {Rees}}{{Efstathiou} \&
  {Rees}}{1988}]{efstathiou1988}
{Efstathiou} G.,  {Rees} M.~J.,  1988, \mnras, 230, 5P

\bibitem[\protect\citeauthoryear{{Elvis} et~al.,}{{Elvis}
  et~al.}{1994}]{elvis1994}
{Elvis} M.,  et~al., 1994, \apjs, 95, 1

\bibitem[\protect\citeauthoryear{{Enoki}, {Nagashima} \& {Gouda}}{{Enoki}
  et~al.}{2003}]{enoki2003}
{Enoki} M.,  {Nagashima} M.,    {Gouda} N.,  2003, PASJ, 55, 133

\bibitem[\protect\citeauthoryear{{Fabian} \& {Nulsen}}{{Fabian} \&
  {Nulsen}}{1977}]{fabian1977}
{Fabian} A.~C.,  {Nulsen} P.~E.~J.,  1977, \mnras, 180, 479

\bibitem[\protect\citeauthoryear{{Fabian}, {Sanders}, {Allen}, {Crawford},
  {Iwasawa}, {Johnstone}, {Schmidt} \& {Taylor}}{{Fabian}
  et~al.}{2003}]{fabian2003}
{Fabian} A.~C.,  {Sanders} J.~S.,  {Allen} S.~W.,  {Crawford} C.~S.,  {Iwasawa}
  K.,  {Johnstone} R.~M.,  {Schmidt} R.~W.,    {Taylor} G.~B.,  2003, \mnras,
  344, L43

\bibitem[\protect\citeauthoryear{{Fan} et~al.,}{{Fan}  et~al.}{2001}]{fan2001a}
{Fan} X.,  et~al., 2001, \aj, 121, 54

\bibitem[\protect\citeauthoryear{{Feoli} \& {Mele}}{{Feoli} \&
  {Mele}}{2007}]{feoli2007}
{Feoli} A.,  {Mele} D.,  2007, preprint, astro-ph/0703675

\bibitem[\protect\citeauthoryear{{Ferrarese}}{{Ferrarese}}{2002}]{ferrarese200%
2}
{Ferrarese} L.,  2002, \apj, 578, 90

\bibitem[\protect\citeauthoryear{{Ferrarese} \& {Ford}}{{Ferrarese} \&
  {Ford}}{2005}]{ferrarese2005}
{Ferrarese} L.,  {Ford} H.,  2005, Space Science Reviews, 116, 523

\bibitem[\protect\citeauthoryear{{Ferrarese} \& {Merritt}}{{Ferrarese} \&
  {Merritt}}{2000}]{ferrarese2000}
{Ferrarese} L.,  {Merritt} D.,  2000, \apjl, 539, L9

\bibitem[\protect\citeauthoryear{{Fontanot}, {Cristiani}, {Monaco}, {Nonino},
  {Vanzella}, {Brandt}, {Grazian} \& {Mao}}{{Fontanot}
  et~al.}{2007}]{fontanot2007}
{Fontanot} F.,  {Cristiani} S.,  {Monaco} P.,  {Nonino} M.,  {Vanzella} E.,
  {Brandt} W.~N.,  {Grazian} A.,    {Mao} J.,  2007, \aap, 461, 39

\bibitem[\protect\citeauthoryear{{Fontanot}, {Monaco}, {Cristiani} \&
  {Tozzi}}{{Fontanot} et~al.}{2006}]{fontanot2006}
{Fontanot} F.,  {Monaco} P.,  {Cristiani} S.,    {Tozzi} P.,  2006, \mnras,
  373, 1173

\bibitem[\protect\citeauthoryear{{Gebhardt} et~al.,}{{Gebhardt}
  et~al.}{2000}]{gebhardt2000}
{Gebhardt} K.,  et~al., 2000, \apjl, 539, L13

\bibitem[\protect\citeauthoryear{{Gilli}, {Comastri} \& {Hasinger}}{{Gilli}
  et~al.}{2007}]{gilli2007}
{Gilli} R.,  {Comastri} A.,    {Hasinger} G.,  2007, \aap, 463, 79

\bibitem[\protect\citeauthoryear{{Gnedin}}{{Gnedin}}{2000}]{gnedin2000}
{Gnedin} N.~Y.,  2000, \apj, 542, 535

\bibitem[\protect\citeauthoryear{{Graham} \& {Driver}}{{Graham} \&
  {Driver}}{2007}]{graham2007}
{Graham} A.~W.,  {Driver} S.~P.,  2007, \apj, 655, 77

\bibitem[\protect\citeauthoryear{{Graham}, {Erwin}, {Caon} \&
  {Trujillo}}{{Graham} et~al.}{2001}]{graham2001}
{Graham} A.~W.,  {Erwin} P.,  {Caon} N.,    {Trujillo} I.,  2001, \apjl, 563,
  L11

\bibitem[\protect\citeauthoryear{{Granato} et~al.,}{{Granato}
  et~al.}{2004}]{granato2004}
{Granato} G.~L.,  et~al., 2004, \apj, 600, 580

\bibitem[\protect\citeauthoryear{{Grazian}, {Cristiani}, {D'Odorico},
  {Omizzolo} \& {Pizzella}}{{Grazian} et~al.}{2000}]{grazian2000}
{Grazian} A.,  {Cristiani} S.,  {D'Odorico} V.,  {Omizzolo} A.,    {Pizzella}
  A.,  2000, \aj, 119, 2540

\bibitem[\protect\citeauthoryear{{Haehnelt} \& {Rees}}{{Haehnelt} \&
  {Rees}}{1993}]{haehnelt1993}
{Haehnelt} M.~G.,  {Rees} M.~J.,  1993, \mnras, 263, 168

\bibitem[\protect\citeauthoryear{{Haiman} \& {Loeb}}{{Haiman} \&
  {Loeb}}{1998}]{haiman1998}
{Haiman} Z.,  {Loeb} A.,  1998, \apj, 503, 505

\bibitem[\protect\citeauthoryear{{Haiman} \& {Menou}}{{Haiman} \&
  {Menou}}{2000}]{haiman2000}
{Haiman} Z.,  {Menou} K.,  2000, \apj, 531, 42

\bibitem[\protect\citeauthoryear{{Hao} et~al.,}{{Hao}  et~al.}{2005}]{hao2005}
{Hao} L.,  et~al., 2005, \aj, 129, 1795

\bibitem[\protect\citeauthoryear{{H{\"a}ring} \& {Rix}}{{H{\"a}ring} \&
  {Rix}}{2004}]{haring2004}
{H{\"a}ring} N.,  {Rix} H.-W.,  2004, \apjl, 604, L89

\bibitem[\protect\citeauthoryear{{Hasinger}, {Miyaji} \& {Schmidt}}{{Hasinger}
  et~al.}{2005}]{hasinger2005}
{Hasinger} G.,  {Miyaji} T.,    {Schmidt} M.,  2005, \aap, 441, 417

\bibitem[\protect\citeauthoryear{{Hatziminaoglou}, {Mathez}, {Solanes},
  {Manrique} \& {Salvador-Sol{\'e}}}{{Hatziminaoglou}
  et~al.}{2003}]{hatziminaoglou2003}
{Hatziminaoglou} E.,  {Mathez} G.,  {Solanes} J.-M.,  {Manrique} A.,
  {Salvador-Sol{\'e}} E.,  2003, \mnras, 343, 692

\bibitem[\protect\citeauthoryear{{Hoeft}, {Yepes}, {Gottl{\"o}ber} \&
  {Springel}}{{Hoeft} et~al.}{2006}]{hoeft2006}
{Hoeft} M.,  {Yepes} G.,  {Gottl{\"o}ber} S.,    {Springel} V.,  2006, \mnras,
  371, 401

\bibitem[\protect\citeauthoryear{{Hopkins}, {Hernquist}, {Cox} \&
  {Keres}}{{Hopkins} et~al.}{2007b}]{hopkins2007b}
{Hopkins} P.~F.,  {Hernquist} L.,  {Cox} T.~J.,    {Keres} D.,  2007b, ArXiv
  e-prints, 706

\bibitem[\protect\citeauthoryear{{Hopkins}, {Hernquist}, {Cox}, {Robertson},
  {Di Matteo} \& {Springel}}{{Hopkins} et~al.}{2006}]{hopkins2006}
{Hopkins} P.~F.,  {Hernquist} L.,  {Cox} T.~J.,  {Robertson} B.,  {Di Matteo}
  T.,    {Springel} V.,  2006, \apj, 639, 700

\bibitem[\protect\citeauthoryear{{Hopkins}, {Hernquist}, {Cox}, {Robertson} \&
  {Krause}}{{Hopkins} et~al.}{2007a}]{hopkins2007a}
{Hopkins} P.~F.,  {Hernquist} L.,  {Cox} T.~J.,  {Robertson} B.,    {Krause}
  E.,  2007a, preprint, astro-ph/0701351

\bibitem[\protect\citeauthoryear{{Hopkins}, {Hernquist}, {Martini}, {Cox},
  {Robertson}, {Di Matteo} \& {Springel}}{{Hopkins} et~al.}{2005}]{hopkins2005}
{Hopkins} P.~F.,  {Hernquist} L.,  {Martini} P.,  {Cox} T.~J.,  {Robertson} B.,
   {Di Matteo} T.,    {Springel} V.,  2005, \apjl, 625, L71

\bibitem[\protect\citeauthoryear{{Hopkins}, {Richards} \&
  {Hernquist}}{{Hopkins} et~al.}{2007}]{hopkins2007}
{Hopkins} P.~F.,  {Richards} G.~T.,    {Hernquist} L.,  2007, \apj, 654, 731

\bibitem[\protect\citeauthoryear{{Hunt}, {Steidel}, {Adelberger} \&
  {Shapley}}{{Hunt} et~al.}{2004}]{hunt2004}
{Hunt} M.~P.,  {Steidel} C.~C.,  {Adelberger} K.~L.,    {Shapley} A.~E.,  2004,
  \apj, 605, 625

\bibitem[\protect\citeauthoryear{{Kauffmann}}{{Kauffmann}}{1996}]{kauffmann199%
6}
{Kauffmann} G.,  1996, \mnras, 281, 475

\bibitem[\protect\citeauthoryear{{Kauffmann} \& {Haehnelt}}{{Kauffmann} \&
  {Haehnelt}}{2000}]{kauffmann2000}
{Kauffmann} G.,  {Haehnelt} M.,  2000, \mnras, 311, 576

\bibitem[\protect\citeauthoryear{{Kennefick}, {Djorgovski} \& {de
  Carvalho}}{{Kennefick} et~al.}{1995}]{kennefick1995}
{Kennefick} J.~D.,  {Djorgovski} S.~G.,    {de Carvalho} R.~R.,  1995, \aj,
  110, 2553

\bibitem[\protect\citeauthoryear{{Kennicutt}
  Jr.}{{Kennicutt}}{1998}]{kennicutt1998}
{Kennicutt} Jr. R.~C.,  1998, \apj, 498, 541

\bibitem[\protect\citeauthoryear{{Kitzbichler} \& {White}}{{Kitzbichler} \&
  {White}}{2007}]{kitzbichler2007}
{Kitzbichler} M.~G.,  {White} S.~D.~M.,  2007, \mnras, 376, 2

\bibitem[\protect\citeauthoryear{{Koehler}, {Groote}, {Reimers} \&
  {Wisotzki}}{{Koehler} et~al.}{1997}]{kohler1997}
{Koehler} T.,  {Groote} D.,  {Reimers} D.,    {Wisotzki} L.,  1997, \aap, 325,
  502

\bibitem[\protect\citeauthoryear{{Kormendy} \& {Richstone}}{{Kormendy} \&
  {Richstone}}{1995}]{kormendy1995}
{Kormendy} J.,  {Richstone} D.,  1995, \araa, 33, 581

\bibitem[\protect\citeauthoryear{{Kravtsov}, {Gnedin} \& {Klypin}}{{Kravtsov}
  et~al.}{2004}]{Kravtsov2004}
{Kravtsov} A.~V.,  {Gnedin} O.~Y.,    {Klypin} A.~A.,  2004, \apj, 609, 482

\bibitem[\protect\citeauthoryear{{La Franca} et~al.,}{{La Franca}
  et~al.}{2005}]{lafranca2005}
{La Franca} F.,  et~al., 2005, \apj, 635, 864

\bibitem[\protect\citeauthoryear{{Lamastra}, {Perola} \& {Matt}}{{Lamastra}
  et~al.}{2006}]{lamastra2006}
{Lamastra} A.,  {Perola} G.~C.,    {Matt} G.,  2006, \aap, 449, 551

\bibitem[\protect\citeauthoryear{{Lemson} \& {Virgo Consortium}}{{Lemson} \&
  {Virgo Consortium}}{2006}]{lemson2006}
{Lemson} G.,  {Virgo Consortium} t.,  2006, preprint, astro-ph/060801

\bibitem[\protect\citeauthoryear{{Li}, {Hernquist}, {Robertson}, {Cox},
  {Hopkins}, {Springel}, {Gao}, {Di Matteo}, {Zentner}, {Jenkins} \&
  {Yoshida}}{{Li} et~al.}{2007}]{li2007}
{Li} Y.,  {Hernquist} L.,  {Robertson} B.,  {Cox} T.~J.,  {Hopkins} P.~F.,
  {Springel} V.,  {Gao} L.,  {Di Matteo} T.,  {Zentner} A.~R.,  {Jenkins} A.,
   {Yoshida} N.,  2007, \apj, 665, 187

\bibitem[\protect\citeauthoryear{{Lidz}, {Hopkins}, {Cox}, {Hernquist} \&
  {Robertson}}{{Lidz} et~al.}{2006}]{lidz2006}
{Lidz} A.,  {Hopkins} P.~F.,  {Cox} T.~J.,  {Hernquist} L.,    {Robertson} B.,
  2006, \apj, 641, 41

\bibitem[\protect\citeauthoryear{{Lynden-Bell}}{{Lynden-Bell}}{1969}]{lynden19%
69}
{Lynden-Bell} D.,  1969, \nat, 223, 690

\bibitem[\protect\citeauthoryear{{Magorrian} et~al.,}{{Magorrian}
  et~al.}{1998}]{magorrian1998}
{Magorrian} J.,  et~al., 1998, \aj, 115, 2285

\bibitem[\protect\citeauthoryear{{Maio}, {Dolag}, {Ciardi} \&
  {Tornatore}}{{Maio} et~al.}{2007}]{maio2007}
{Maio} U.,  {Dolag} K.,  {Ciardi} B.,    {Tornatore} L.,  2007, \mnras, 379,
  963

\bibitem[\protect\citeauthoryear{{Malbon}, {Baugh}, {Frenk} \&
  {Lacey}}{{Malbon} et~al.}{2006}]{malbon2006}
{Malbon} R.~K.,  {Baugh} C.~M.,  {Frenk} C.~S.,    {Lacey} C.~G.,  2006,
  preprint, astro-ph/0607424

\bibitem[\protect\citeauthoryear{{Marconi}, {Risaliti}, {Gilli}, {Hunt},
  {Maiolino} \& {Salvati}}{{Marconi} et~al.}{2004}]{marconi2004}
{Marconi} A.,  {Risaliti} G.,  {Gilli} R.,  {Hunt} L.~K.,  {Maiolino} R.,
  {Salvati} M.,  2004, \mnras, 351, 169

\bibitem[\protect\citeauthoryear{{Martini} \& {Weinberg}}{{Martini} \&
  {Weinberg}}{2001}]{martini2001}
{Martini} P.,  {Weinberg} D.~H.,  2001, \apj, 547, 12

\bibitem[\protect\citeauthoryear{{Marulli}, {Branchini}, {Moscardini} \&
  {Volonteri}}{{Marulli} et~al.}{2007}]{marulli2007}
{Marulli} F.,  {Branchini} E.,  {Moscardini} L.,    {Volonteri} M.,  2007,
  \mnras, 375, 649

\bibitem[\protect\citeauthoryear{{Marulli}, {Crociani}, {Volonteri},
  {Branchini} \& {Moscardini}}{{Marulli} et~al.}{2006}]{marulli2006}
{Marulli} F.,  {Crociani} D.,  {Volonteri} M.,  {Branchini} E.,    {Moscardini}
  L.,  2006, \mnras, 368, 1269

\bibitem[\protect\citeauthoryear{{Matute}, {La Franca}, {Pozzi}, {Gruppioni},
  {Lari} \& {Zamorani}}{{Matute} et~al.}{2006}]{matute2006}
{Matute} I.,  {La Franca} F.,  {Pozzi} F.,  {Gruppioni} C.,  {Lari} C.,
  {Zamorani} G.,  2006, \aap, 451, 443

\bibitem[\protect\citeauthoryear{{McLure} \& {Dunlop}}{{McLure} \&
  {Dunlop}}{2002}]{mclure2002}
{McLure} R.~J.,  {Dunlop} J.~S.,  2002, \mnras, 331, 795

\bibitem[\protect\citeauthoryear{{McNamara}, {Nulsen}, {Wise}, {Rafferty},
  {Carilli}, {Sarazin} \& {Blanton}}{{McNamara} et~al.}{2005}]{mcnamara2005}
{McNamara} B.~R.,  {Nulsen} P.~E.~J.,  {Wise} M.~W.,  {Rafferty} D.~A.,
  {Carilli} C.,  {Sarazin} C.~L.,    {Blanton} E.~L.,  2005, \nat, 433, 45

\bibitem[\protect\citeauthoryear{{Miyaji}, {Hasinger} \& {Schmidt}}{{Miyaji}
  et~al.}{2000}]{miyaji2000}
{Miyaji} T.,  {Hasinger} G.,    {Schmidt} M.,  2000, \aap, 353, 25

\bibitem[\protect\citeauthoryear{{Miyaji}, {Hasinger} \& {Schmidt}}{{Miyaji}
  et~al.}{2001}]{miyaji2001}
{Miyaji} T.,  {Hasinger} G.,    {Schmidt} M.,  2001, \aap, 369, 49

\bibitem[\protect\citeauthoryear{{Mo}, {Mao} \& {White}}{{Mo}
  et~al.}{1998}]{mo1998}
{Mo} H.~J.,  {Mao} S.,    {White} S.~D.~M.,  1998, \mnras, 295, 319

\bibitem[\protect\citeauthoryear{{Morandi} \& {Ettori}}{{Morandi} \&
  {Ettori}}{2007}]{morandi2007}
{Morandi} A.,  {Ettori} S.,  2007, \mnras, pp 774--+

\bibitem[\protect\citeauthoryear{{Nagar}, {Falcke} \& {Wilson}}{{Nagar}
  et~al.}{2005}]{nagar2005}
{Nagar} N.~M.,  {Falcke} H.,    {Wilson} A.~S.,  2005, \aap, 435, 521

\bibitem[\protect\citeauthoryear{{Nandra}, {Laird} \& {Steidel}}{{Nandra}
  et~al.}{2005}]{nandra2005}
{Nandra} K.,  {Laird} E.~S.,    {Steidel} C.~C.,  2005, \mnras, 360, L39

\bibitem[\protect\citeauthoryear{{Netzer} \& {Trakhtenbrot}}{{Netzer} \&
  {Trakhtenbrot}}{2007}]{netzer2007}
{Netzer} H.,  {Trakhtenbrot} B.,  2007, \apj, 654, 754

\bibitem[\protect\citeauthoryear{{Novak}, {Faber} \& {Dekel}}{{Novak}
  et~al.}{2006}]{novak2006}
{Novak} G.~S.,  {Faber} S.~M.,    {Dekel} A.,  2006, \apj, 637, 96

\bibitem[\protect\citeauthoryear{{Percival} \& {Miller}}{{Percival} \&
  {Miller}}{1999}]{percival1999}
{Percival} W.,  {Miller} L.,  1999, \mnras, 309, 823

\bibitem[\protect\citeauthoryear{{Peterson}, {Paerels}, {Kaastra}, {Arnaud},
  {Reiprich}, {Fabian}, {Mushotzky}, {Jernigan} \& {Sakelliou}}{{Peterson}
  et~al.}{2001}]{peterson2001}
{Peterson} J.~R.,  {Paerels} F.~B.~S.,  {Kaastra} J.~S.,  {Arnaud} M.,
  {Reiprich} T.~H.,  {Fabian} A.~C.,  {Mushotzky} R.~F.,  {Jernigan} J.~G.,
  {Sakelliou} I.,  2001, \aap, 365, L104

\bibitem[\protect\citeauthoryear{{Richards} et~al.,}{{Richards}
  et~al.}{2005}]{richards2005}
{Richards} G.~T.,  et~al., 2005, \mnras, 360, 839

\bibitem[\protect\citeauthoryear{{Richards} et~al.,}{{Richards}
  et~al.}{2006}]{richards2006}
{Richards} G.~T.,  et~al., 2006, \aj, 131, 2766

\bibitem[\protect\citeauthoryear{{Richstone} et~al.,}{{Richstone}
  et~al.}{1998}]{richstone1998}
{Richstone} D.,  et~al., 1998, \nat, 395, A14+

\bibitem[\protect\citeauthoryear{{Salpeter}}{{Salpeter}}{1964}]{salpeter1964}
{Salpeter} E.~E.,  1964, \apj, 140, 796

\bibitem[\protect\citeauthoryear{{S{\'a}nchez}, {Baugh}, {Percival}, {Peacock},
  {Padilla}, {Cole}, {Frenk} \& {Norberg}}{{S{\'a}nchez}
  et~al.}{2006}]{Sanchez2006}
{S{\'a}nchez} A.~G.,  {Baugh} C.~M.,  {Percival} W.~J.,  {Peacock} J.~A.,
  {Padilla} N.~D.,  {Cole} S.,  {Frenk} C.~S.,    {Norberg} P.,  2006, \mnras,
  366, 189

\bibitem[\protect\citeauthoryear{{Sazonov} \& {Revnivtsev}}{{Sazonov} \&
  {Revnivtsev}}{2004}]{sazonov2004}
{Sazonov} S.~Y.,  {Revnivtsev} M.~G.,  2004, \aap, 423, 469

\bibitem[\protect\citeauthoryear{{Schmidt}, {Schneider} \& {Gunn}}{{Schmidt}
  et~al.}{1995}]{schmidt1995}
{Schmidt} M.,  {Schneider} D.~P.,    {Gunn} J.~E.,  1995, \aj, 110, 68

\bibitem[\protect\citeauthoryear{{Seljak}}{{Seljak}}{2002}]{seljak2002}
{Seljak} U.,  2002, \mnras, 334, 797

\bibitem[\protect\citeauthoryear{{Shankar} \& {Mathur}}{{Shankar} \&
  {Mathur}}{2007}]{shankar2007}
{Shankar} F.,  {Mathur} S.,  2007, \apj, 660, 1051

\bibitem[\protect\citeauthoryear{{Shankar}, {Salucci}, {Granato}, {De Zotti} \&
  {Danese}}{{Shankar} et~al.}{2004}]{shankar2004}
{Shankar} F.,  {Salucci} P.,  {Granato} G.~L.,  {De Zotti} G.,    {Danese} L.,
  2004, \mnras, 354, 1020

\bibitem[\protect\citeauthoryear{{Shinozaki}, {Miyaji}, {Ishisaki}, {Ueda} \&
  {Ogasaka}}{{Shinozaki} et~al.}{2006}]{shinozaki2006}
{Shinozaki} K.,  {Miyaji} T.,  {Ishisaki} Y.,  {Ueda} Y.,    {Ogasaka} Y.,
  2006, \aj, 131, 2843

\bibitem[\protect\citeauthoryear{{Siana} et~al.,}{{Siana}
  et~al.}{2006}]{siana2006}
{Siana} B.,  et~al., 2006, preprint, astro-ph/0604373

\bibitem[\protect\citeauthoryear{{Sijacki}, {Springel}, {di Matteo} \&
  {Hernquist}}{{Sijacki} et~al.}{2007}]{sijacki2007}
{Sijacki} D.,  {Springel} V.,  {di Matteo} T.,    {Hernquist} L.,  2007,
  \mnras, 380, 877

\bibitem[\protect\citeauthoryear{{Silverman} et~al.,}{{Silverman}
  et~al.}{2005a}]{silverman2005a}
{Silverman} J.~D.,  et~al., 2005a, \apj, 618, 123

\bibitem[\protect\citeauthoryear{{Silverman} et~al.,}{{Silverman}
  et~al.}{2005b}]{silverman2005b}
{Silverman} J.~D.,  et~al., 2005b, \apj, 624, 630

\bibitem[\protect\citeauthoryear{{Soltan}}{{Soltan}}{1982}]{soltan1982}
{Soltan} A.,  1982, \mnras, 200, 115

\bibitem[\protect\citeauthoryear{{Somerville}, {Primack} \&
  {Faber}}{{Somerville} et~al.}{2001}]{somerville2001}
{Somerville} R.~S.,  {Primack} J.~R.,    {Faber} S.~M.,  2001, \mnras, 320, 504

\bibitem[\protect\citeauthoryear{{Spergel} et~al.,}{{Spergel}
  et~al.}{2003}]{spergel2003}
{Spergel} D.~N.,  et~al., 2003, \apjs, 148, 175

\bibitem[\protect\citeauthoryear{{Spergel} et~al.,}{{Spergel}
  et~al.}{2007}]{spergel2007}
{Spergel} D.~N.,  et~al., 2007, \apjs, 170, 377

\bibitem[\protect\citeauthoryear{{Springel}}{{Springel}}{2005}]{springel2005ga%
dget2}
{Springel} V.,  2005, \mnras, 364, 1105

\bibitem[\protect\citeauthoryear{{Springel}, {Di Matteo} \&
  {Hernquist}}{{Springel} et~al.}{2005}]{springel2005c}
{Springel} V.,  {Di Matteo} T.,    {Hernquist} L.,  2005, \mnras, 361, 776

\bibitem[\protect\citeauthoryear{{Springel} et~al.,}{{Springel}
  et~al.}{2005}]{springel2005}
{Springel} V.,  et~al., 2005, \nat, 435, 629

\bibitem[\protect\citeauthoryear{{Springel}, {White}, {Tormen} \&
  {Kauffmann}}{{Springel} et~al.}{2001}]{springel2001b}
{Springel} V.,  {White} S.~D.~M.,  {Tormen} G.,    {Kauffmann} G.,  2001,
  \mnras, 328, 726

\bibitem[\protect\citeauthoryear{{Springel}, {Yoshida} \& {White}}{{Springel}
  et~al.}{2001}]{springel2001}
{Springel} V.,  {Yoshida} N.,    {White} S.~D.~M.,  2001, New Astronomy, 6, 79

\bibitem[\protect\citeauthoryear{{Sutherland} \& {Dopita}}{{Sutherland} \&
  {Dopita}}{1993}]{sutherland1993}
{Sutherland} R.~S.,  {Dopita} M.~A.,  1993, \apjs, 88, 253

\bibitem[\protect\citeauthoryear{{Tamura}, {Kaastra}, {Peterson}, {Paerels},
  {Mittaz}, {Trudolyubov}, {Stewart}, {Fabian}, {Mushotzky}, {Lumb} \&
  {Ikebe}}{{Tamura} et~al.}{2001}]{tamura2001}
{Tamura} T.,  {Kaastra} J.~S.,  {Peterson} J.~R.,  {Paerels} F.~B.~S.,
  {Mittaz} J.~P.~D.,  {Trudolyubov} S.~P.,  {Stewart} G.,  {Fabian} A.~C.,
  {Mushotzky} R.~F.,  {Lumb} D.~H.,    {Ikebe} Y.,  2001, \aap, 365, L87

\bibitem[\protect\citeauthoryear{{Tremaine} et~al.,}{{Tremaine}
  et~al.}{2002}]{tremaine2002}
{Tremaine} S.,  et~al., 2002, \apj, 574, 740

\bibitem[\protect\citeauthoryear{{Tundo}, {Bernardi}, {Hyde}, {Sheth} \&
  {Pizzella}}{{Tundo} et~al.}{2007}]{tundo2007}
{Tundo} E.,  {Bernardi} M.,  {Hyde} J.~B.,  {Sheth} R.~K.,    {Pizzella} A.,
  2007, \apj, 663, 53

\bibitem[\protect\citeauthoryear{{Ueda}, {Akiyama}, {Ohta} \& {Miyaji}}{{Ueda}
  et~al.}{2003}]{ueda2003}
{Ueda} Y.,  {Akiyama} M.,  {Ohta} K.,    {Miyaji} T.,  2003, \apj, 598, 886

\bibitem[\protect\citeauthoryear{{Viola}, {Monaco}, {Borgani}, {Murante} \&
  {Tornatore}}{{Viola} et~al.}{2007}]{viola2007}
{Viola} M.,  {Monaco} P.,  {Borgani} S.,  {Murante} G.,    {Tornatore} L.,
  2007, ArXiv e-prints, 710

\bibitem[\protect\citeauthoryear{{Volonteri}, {Haardt} \& {Madau}}{{Volonteri}
  et~al.}{2003}]{volonteri2003a}
{Volonteri} M.,  {Haardt} F.,    {Madau} P.,  2003, \apj, 582, 559

\bibitem[\protect\citeauthoryear{{Volonteri}, {Sikora} \& {Lasota}}{{Volonteri}
  et~al.}{2007}]{volonteri2007}
{Volonteri} M.,  {Sikora} M.,    {Lasota} J.-P.,  2007, preprint,
  astro-ph/0706.3900, 706

\bibitem[\protect\citeauthoryear{{Wang}, {De Lucia}, {Kitzbichler} \&
  {White}}{{Wang} et~al.}{2007}]{wang2007}
{Wang} J.,  {De Lucia} G.,  {Kitzbichler} M.~G.,    {White} S.~D.~M.,  2007,
  preprint, astro-ph/0706.2551, 706

\bibitem[\protect\citeauthoryear{{Weinmann}, {van den Bosch}, {Yang}, {Mo},
  {Croton} \& {Moore}}{{Weinmann} et~al.}{2006}]{weinmann2006}
{Weinmann} S.~M.,  {van den Bosch} F.~C.,  {Yang} X.,  {Mo} H.~J.,  {Croton}
  D.~J.,    {Moore} B.,  2006, \mnras, 372, 1161

\bibitem[\protect\citeauthoryear{{White} \& {Frenk}}{{White} \&
  {Frenk}}{1991}]{white1991}
{White} S.~D.~M.,  {Frenk} C.~S.,  1991, \apj, 379, 52

\bibitem[\protect\citeauthoryear{{White} \& {Rees}}{{White} \&
  {Rees}}{1978}]{white1978}
{White} S.~D.~M.,  {Rees} M.~J.,  1978, \mnras, 183, 341

\bibitem[\protect\citeauthoryear{{Wolf}, {Wisotzki}, {Borch}, {Dye},
  {Kleinheinrich} \& {Meisenheimer}}{{Wolf} et~al.}{2003}]{wolf2003}
{Wolf} C.,  {Wisotzki} L.,  {Borch} A.,  {Dye} S.,  {Kleinheinrich} M.,
  {Meisenheimer} K.,  2003, \aap, 408, 499

\bibitem[\protect\citeauthoryear{{Wyithe}}{{Wyithe}}{2006}]{wyithe2006}
{Wyithe} J.~S.~B.,  2006, \mnras, 365, 1082

\bibitem[\protect\citeauthoryear{{Wyithe} \& {Loeb}}{{Wyithe} \&
  {Loeb}}{2003}]{wyithe2003}
{Wyithe} J.~S.~B.,  {Loeb} A.,  2003, \apj, 595, 614

\end{thebibliography}

\label{lastpage}

\end{document}